\documentclass[12pt]{iopart}
\usepackage{graphicx}
\usepackage{iopams}  
\newcommand{\ket}[1]{\ensuremath{|{#1}\rangle}}

\begin{document}

\title{The effect of core-collapse supernova accretion phase turbulence on neutrino flavor evolution}

\author{James P.\ Kneller}
\address{Department of Physics, North Carolina State University, Raleigh, North Carolina 27695-8202, USA}

\author{Mithi de los Reyes}
\address{Institute of Astronomy, University of Cambridge, Madingley Road, Cambridge CB3 0HA, UK}

\ead{\mailto{jim\_kneller@ncsu.edu} \mailto{macd4@ast.cam.ac.uk} }

\begin{abstract}

During the accretion phase of a core-collapse supernovae, large amplitude turbulence is generated by the combination of the standing accretion shock instability and convection driven by neutrino heating. The turbulence directly affects the dynamics of the explosion, but there is also the possibility of an additional, indirect, feedback mechanism due to the effect turbulence can have upon neutrino flavor evolution and thus the neutrino heating. In this paper we consider the effect of turbulence during the accretion phase upon neutrino evolution, both numerically and analytically. Adopting representative supernova profiles taken from the accretion phase of a supernova simulation, we find the numerical calculations exhibit no effect from turbulence. We explain this absence using two analytic descriptions: the Stimulated Transition model and the Distorted Phase Effect model. In the Stimulated Transition model turbulence effects depend upon six different lengthscales, and three criteria must be satisfied between them if one is to observe a change in the flavor evolution due to Stimulated Transition. We further demonstrate that the Distorted Phase Effect depends upon the presence of multiple semi-adiabatic MSW resonances or discontinuities that also can be expressed as a relationship between three of the same lengthscales. When we examine the supernova profiles used in the numerical calculations we find the three Stimulated Transition criteria cannot be satisfied, independent of the form of the turbulence power spectrum, and that the same supernova profiles lack the multiple semi-adiabatic MSW resonances or discontinuities necessary to produce a Distorted Phase Effect. Thus we conclude that even though large amplitude turbulence is present in supernova during the accretion phase, it has no effect upon neutrino flavor evolution.

\end{abstract}

\maketitle

\section{Introduction}

Modeling the collapse of the core of a supermassive star has advanced by huge strides in recent years. Spherically symmetric simulations, including the best microphysics available, do not explode \cite{2001PhRvD..63j3004L,2001PhRvL..86.1935M}, indicating phenomena which only occur in two and three spatial dimensions are crucial for the explosion. When simulations are undertaken with extra dimensions, new features emerge such as neutrino-driven convection and the Standing Accretion Shock Instability \cite{2003ApJ...584..971B,2006A&A...447.1049B,2006A&A...453..661K,2007PhR...442...23B,2008ApJ...678.1207I,2013ApJ...775...35C,2014ApJ...785..123C}. Both these processes generate large fluid motions during the first second of the explosion, breaking spherical symmetry. Thus well-studied aspherical features of observed core-collapse supernovae -- such as the high-velocity ``jets'' of sulfur-rich material seen in the supernova remnant Cassiopeia A \cite{2013ApJ...772..134M}, the double-peaked structure of the Oxygen and Magnesium nebular lines in observations of SN 2003jd \cite{2005Sci...308.1284M}, the two light curve components of SN 2013ge \cite{2016ApJ...821...57D}, and the spectropolarimetric observations of stripped-envelope core-collapse supernovae \cite{2012ApJ...754...63T} -- are apparently generated during the earliest moments of the explosion. 

The violent fluid motions in the core of the star naturally lead to turbulence and the role of the turbulence in the dynamics of the explosion has become a topic of recent interest \cite{2011ApJ...742...74M,2012ApJ...749..142F,2013ApJ...765..110D,2013PhST..155a4022E,2014ApJ...783..125H,2015ApJ...808...70A,2015ApJ...799....5C,2015ApJ...808...70A,2016ApJ...820...76R}. At present, core-collapse supernova simulations are beginning to reach sufficient spatial resolution to see the inertial range of the turbulence \cite{2016ApJ...820...76R}, which will allow simulators to better answer questions such as the effect of the `turbulence bottleneck' -- the accumulation of power in the inertial scales of the turbulence \cite{1994PhFl....6.1411F,2003PhRvE..68b6304D,2015PhRvE..92c3009K,2015ApJ...808...70A} -- and the difference between turbulence in 2D and 3D. However, the presence of turbulence in the simulations also raises new questions, especially regarding the neutrinos which play a crucial role in the explosion. Several studies have shown that turbulence during the \emph{cooling} phase of the explosion changes the neutrino flavor evolution  \cite{Sawyer:1990PhRvD..42.3908S,Loreti:1995ae,1996PhRvD..54.3941B,Friedland:2006ta,Fogli:2006JCAP...06..012F,Choubey:2007PhRvD..76g3013C,2010PhRvD..82l3004K,2013PhRvD..88d5020K,2013PhRvD..88b5004K,2013PhRvD..88b3008L,2015PhRvD..92a3009K}. Given this result, one might also expect turbulence during the \emph{accretion} phase to also have an effect upon the neutrino flavor evolution. Any change in the flavor evolution will alter the neutrino spectra which would then alter the amount of neutrino heating. Since neutrino heating is a source of the fluid motion which leads to turbulence, any effect of turbulence upon the neutrinos feeds back to the turbulence itself. 

A previous study by Reid, Adams, \& Seunarine considered the effect of accretion phase turbulence but found no effect upon the neutrino flavor evolution \cite{2011PhRvD..84h5023R}. However, this study was purely numerical and introduced a large number of parameters (e.g., the density profile for the supernova, the turbulence power spectrum). Scanning the full multi-dimensional parameter space is the only way to conclusively determine if some combination of parameters might produce flavor evolution. This is not feasible, particularly given the large dynamic range inherent in numerical calculations of the effect of turbulence upon neutrino evolution, which makes these calculations extremely time-consuming. Perhaps surprisingly, the effect of turbulence upon neutrino flavor evolution is actually amenable to theory even on a case-by-case basis \cite{Kneller:2012id,Patton:2013dba,Patton:2014lza}. Armed with analytical tools, one can more efficiently explore the parameter space and can make confident statements about the effect of turbulence upon the neutrinos, despite the many uncertainties that still exist. The goal of this paper is to use the analytical tools describing the effect of turbulence upon neutrino flavor evolution to explore why no effect of turbulence was  seen in the study by Reid, Adams \& Seunarine \cite{2011PhRvD..84h5023R} and whether there is any corner of the parameter space where it might occur. 

Our paper is organized as follows. In Section \S\ref{sec:neutrinopropagation} we present the general description for neutrino propagation through matter, describe how we model the turbulence during the accretion phase, and then demonstrate using numerical calculations that no significant effect of turbulence occurs for the neutrinos using a set of representative density profiles. We then examine the problem using analytic models of the effect of turbulence on neutrino flavor evolution; namely, the Stimulated Transition and Distorted Phase Effect models. In Section \S\ref{sec:thescales} we show the Stimulated Transition model depends upon six different lengthscales and argue that three criteria must be satisfied if turbulence is to have an effect via this mechanism. When we assess whether these criteria are met in the numerical examples presented in Section \S\ref{sec:neutrinopropagation}, we see that they fail by many orders of magnitude. Section \S\ref{sec:distortedphase} considers whether turbulence effects should be expected via the Distorted Phase Effect, which depends upon the presence of multiple discontinuities or semi-adiabatic MSW resonances. When we search the profiles used in the numerical examples we find there are none. We then make our conclusions in Section \S\ref{sec:Conclusions}, discuss the limitations of our study, and determine whether one should ever expect turbulence effects upon the neutrinos during the accretion phase of core-collapse supernovae. 


\section{Neutrino Propagation}
\label{sec:neutrinopropagation}

The flavour evolution of a neutrino can be described in terms of elementary quantum mechanics. Since the energy of the neutrino is typically much larger than its mass, the velocity of the neutrino is very close to the speed of light, allowing us to relate time evolution to spatial evolution. At some initial position $r_0$, the neutrino flavor state is given by $\ket{\phi(r_0)}$. The neutrino propagates to a point $r$, where its state is given by $\ket{\phi(r)}$. The states at the two positions are related via the evolution matrix $S(r,r_0)$, which is the solution of the Schr\"odinger equation 
\begin{equation}
 \rmi \frac{dS}{dr} = H S,
\end{equation}
where $H$ is the Hamiltonian. The same equation applies for antineutrinos using an evolution matrix $\bar{S}$ and a Hamiltonian $\bar{H}$. The Hamiltonian is composed of multiple terms: the vacuum, $H_V$, the matter potential, $H_M$, and the neutrino self-interaction $H_{SI}$. For the purposes of this paper we shall ignore the contribution from $H_{SI}$ based on the result from Chakraborty \emph{et al.} \cite{2011PhRvL.107o1101C}, which indicates no effect from self-interactions during the accretion phase. The vacuum Hamiltonian $H_{V}$ is diagonal in the mass basis (which is denoted by the superscript $(m)$) and given by
\begin{equation}
 H^{(m)}_V = \frac{1}{2E}\left( \begin{array}{*{20}{c}} m_1^2 & 0 & 0  \\ 0 & m_2^2 & 0  \\ 0 & 0 & m_3^2 \end{array} \right).
\end{equation}
In this equation $E$ is the neutrino energy and $m_i$ are the neutrino masses. 
The vacuum Hamiltonian for the antineutrinos, $\bar{H}^{(m)}_V$, is the same as for the neutrinos.
The vacuum Hamiltonian in the flavor basis (which is denoted by the superscript $(f)$) is related to the vacuum Hamiltonian in the mass basis via a unitary matrix $U$ such that $H_V^{(f)} = U_V H^{(m)}_V U_V^{\dag}$. The mixing matrix $U$ can be built from the product of a set of complex rotation matrices. We define a matrix $R_{ij}(\theta_{ij},\delta_{ij})$ such that elements $ii$ and $jj$ are equal to $\cos\theta_{ij}$, element $ij$ is $e^{\rmi\delta_{ij}}\sin\theta_{ij}$ and element $ji$ is the negative complex conjugate of $ij$. All other off-diagonal elements of $R_{ij}$ are zero and all other diagonal elements are set to unity. Using these matrices, $U_V$ can be expressed as 
\begin{equation}
 U_V = R_{23}(\theta_{23},0)\,R_{13}(\theta_{13},\delta_{13})\,R_{12}(\theta_{12},0).
 \end{equation}
We have omitted the additional phases which can be included in the definition of $U_V$ because it has been shown they do not affect flavor oscillations \cite{1987NuPhB.282..589L,2009PhRvD..80e3002K,2012JPhG...39c5201G}.
The equivalent mixing matrix for the antineutrinos, $\bar{U}_V$, is the complex conjugate of ${U}_V$.

To the vacuum Hamiltonian we must add the effect of matter. For three active flavors of neutrino, it is enough to consider just the Charged Current potential $V^{\,C}$ of the electron flavor neutrinos
\begin{equation}
 V^{\,C} = \sqrt{2}\,G_F\,n_e
 \label{eq:VC}
\end{equation}
because the Neutral Current potential is the same for each flavor, leading to an overall common phase shift which is unobservable. Here $G_F$ is Fermi's constant and $n_e$ is the electron density which is equal to $n_e = \rho Y_e / m_N$ where $\rho$ is the mass density, $Y_e$ the electron fraction and $m_N$ the nucleon mass. Using the ordering $e,\mu,\tau$ for the rows and columns of matrices in the flavor basis, the matter potential in the flavor basis $H^{(f)}_M$ is 
\begin{equation}
H^{(f)}_M = \left( \begin{array}{*{20}{c}} V^{\,C} & 0 & 0  \\ 0 & 0 & 0  \\ 0 & 0 &0 \end{array} \right).
\end{equation}
The matter Hamiltonian for the antineutrinos, $\bar{H}_M$, is simply $\bar{H}_M = - H_M$.

Cases where the electron density is a smooth function of position, $r$, lead to the phenomenon of neutrino resonances named after Mikheyev, Smirnov and Wolfenstein (MSW) \cite{1978PhRvD..17.2369W,Mikheyev:1985aa,Mikheyev:1986tj}. In dense matter, the flavor states are very similar to the eigenstates of the Hamiltonian. As the mass density drops, the degree of similarity decreases until at a MSW resonance the eigenstates are equal mixtures of two flavor states (and vice versa). The MSW resonance occurs at those locations where the mass density and electron fraction satisfy the equation 
\begin{equation} 
\frac{ m_i^2\,\left|U_{\alpha i}\right|^2 + m_j^2\,\left|U_{\alpha j}\right|^2 }{2E} + H_{M;\alpha\alpha} = \frac{ m_i^2\,\left|U_{\beta i}\right|^2 + m_j^2\,\left|U_{\beta j}\right|^2 }{2E} + H_{M;\beta\beta} \label{eq:MSWresonance}
\end{equation}
where the subscripts $\alpha$, $\beta$ represent generic flavor indices, and $i$ and $j$ are generic mass basis indices. We shall denote the MSW densities by $\rho_{MSW,ij}$.

\begin{figure*}[t!] 
\includegraphics[width=0.49\linewidth]{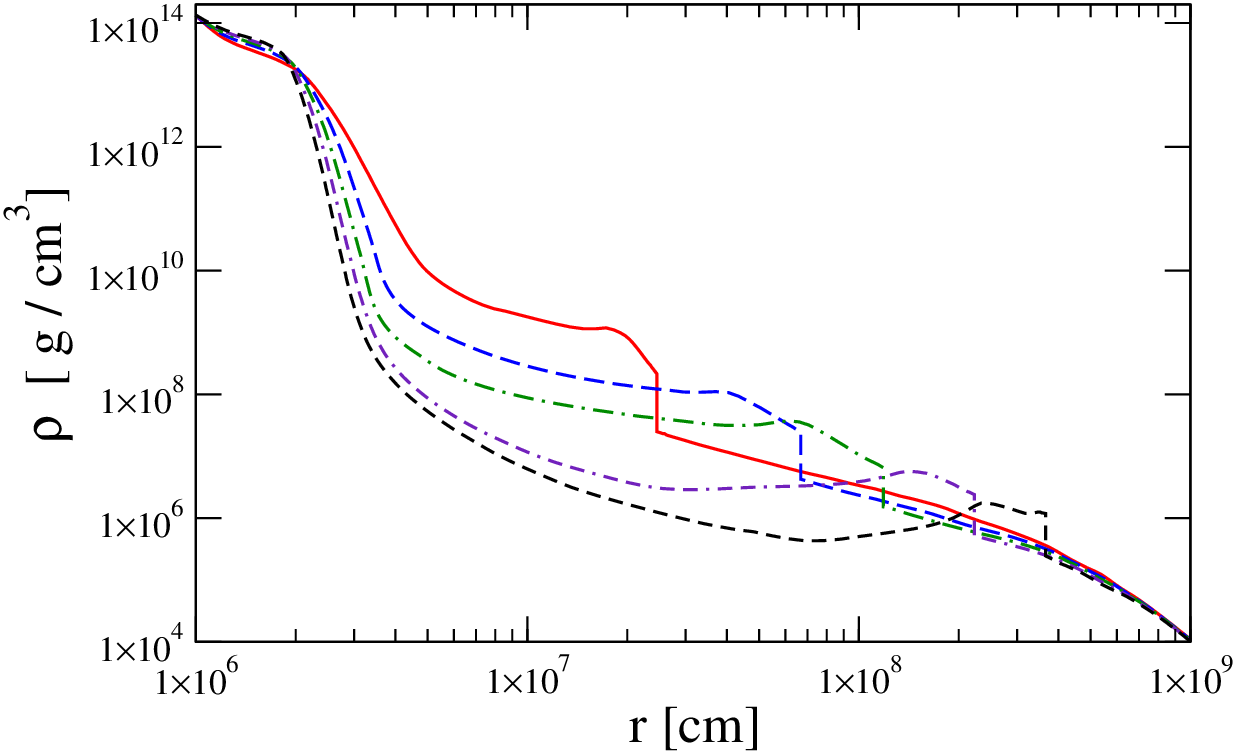}
\includegraphics[width=0.49\linewidth]{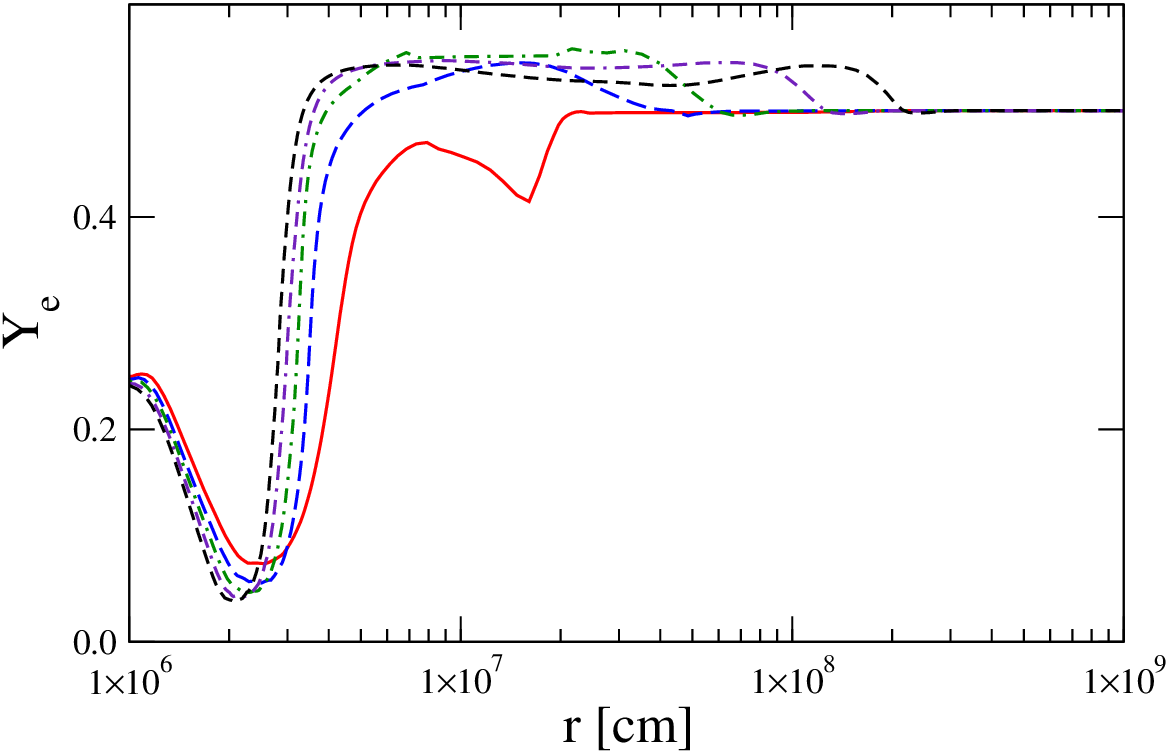}
\caption{The mass density (left panel) and electron fraction (right panel) as a function of distance at various epochs post-bounce from the simulation by Fischer \emph{et al.}  \cite{2010A&A...517A..80F} of the explosion of a $M=10.8\;M_{\odot}$ progenitor. In both panels the epochs are $t=0.3\;{\rm s}$ (solid), $t=0.4\;{\rm s}$ (long dashed), $t=0.5\;{\rm s}$ (dash dot), $t=0.6\;{\rm s}$ (double dash dot), and $t=0.7\;{\rm s}$ (short dashed). \label{fig:rhoYe}}
\end{figure*}

\subsection{Turbulence}

Turbulence enters the Hamiltonian via fluctuations in the electron density $n_e(r)$. Rather than treat the effect of turbulence as a more complicated MSW effect, we choose to treat the neutrino using the models we shall describe in Sections \S\ref{sec:thescales} and \S\ref{sec:distortedphase}. For now, we concentrate on the numerical calculations and the implementation of turbulence.

One might expect that in order to study turbulence numerically, we would adopt mass density profiles and electron fractions from multi-dimensional turbulent supernova simulations, insert these quantities first into equation (\ref{eq:VC}) and then into the Schr\"odinger equation, and begin the numerical integration. For this approach to be successful, several criteria would need to be met by the simulations. First, the simulations need to be three-dimensional, not two-dimensional, because turbulence in two spatial dimensions is found to be quite different from turbulence in three dimensions \cite{2013ApJ...765..110D,2015ApJ...799....5C}. Furthermore, the spatial resolution of the simulation must be at least as small as the oscillation wavelength; in dense matter, this is of order $1/V^{\,C}$. For a mass density of $\rho \sim 10^{13}\;{\rm g/cm^3}$ and $Y_e \sim 0.25$, one estimates -- using the asymptotic formula given in Kneller \& McLaughlin \cite{2009PhRvD..80e3002K} for the eigenvalue splitting at high density -- that the oscillation wavelengths at the neutrinosphere are as small as $10\;{\rm \mu m}$ between some pairs of eigenstates. This lengthscale is many orders of magnitude smaller than the resolution of modern supernova simulations. Thus we find the ideal approach will not work. 

In the absence of profiles from suitable multi-dimensional simulations, we adopt the common alternative approach of using mass density profiles and electron fractions from one dimensional studies and inserting turbulence into them via a model. We use mass density profiles from the hydrodynamical simulation by the Basel group of the explosion of a $10.8\;M_{\odot}$ progenitor \cite{2010A&A...517A..80F}. Several of these profiles are shown in figure~(\ref{fig:rhoYe}). In each profile there appears a shock at radius $R_S$ which we have steepened `by hand' into a discontinuity \cite{2013PhRvD..88b3008L}. During the accretion phase, the shock is approximately stationary; this corresponds to the profile at $t=0.3\;{\rm s}$ in figure~(\ref{fig:rhoYe}). In two- and three-dimensional simulations it is found significant accretion can continue to occur via `downflows' even when the shock begins to advance \cite{2015MNRAS.453..287M,2016ApJ...818..123B}. If these downflows are turbulent, then neutrino evolution along these directions may experience turbulence effects, thus altering the supernova dynamics. In order to investigate this possibility we shall consider profiles at the later snapshot times shown in figure (\ref{fig:rhoYe}) in addition to the standing accretion shock profile. 

Turbulence is inserted into the profile in the region below the shock radius $R_S$ by multiplying the charge current potentials derived from these mass density and electron fraction profiles by a factor of $1 + F(r)$. Here, $F(r)$ is a Gaussian random field which can be represented by a Fourier series
\begin{equation}\label{eq:F1D}
F(r)=C_{\star}\,\Omega(r;R_S)\,\sum_{a=1}^{N_q}\, G_{a} \sin\left(q_{a}\,r + \eta_a\right). 
\end{equation}
The constant $C_{\star}$ sets the overall scale of the turbulence, and the function $\Omega(r;R_S)$ is an `envelope' function that depends upon the shock radius. The sets $\{G\}$, $\{q\}$ and $\{\eta\}$ are random variates generated using some algorithm so as to reproduce the statistical properties of the field, i.e.\ that $F(r)$ has a mean of zero and that the rms of the field is $C_{\star}\,\Omega(r;R_S)$. The power spectrum for the turbulence $E(q)$, where $q$ is the wavenumber, during the accretion phase is uncertain due to the limitations of current simulations. For simplicity we shall adopt an inverse power law 
\begin{equation}
E(q) = \frac{(\alpha-1)}{2\,q_{cut}} \left( \frac{q_{cut}}{|q|}\right)^{\alpha}\, \Theta(|q|-q_{cut}). \label{eq:E1D}
\end{equation}
where $\alpha$ is the spectral index and $q_{cut}$ is the long wavelength / short wavenumber cutoff. $\Theta(|q|-q_{cut})$ is the Heaviside step function and we take $q_{cut}$ to be twice the radius $R_S$ of the shock since modes with longer wavelengths could not be supported, i.e.\ $q_{cut} = \pi/R_S$. 

\begin{figure}[t!] 
\includegraphics[width=\linewidth]{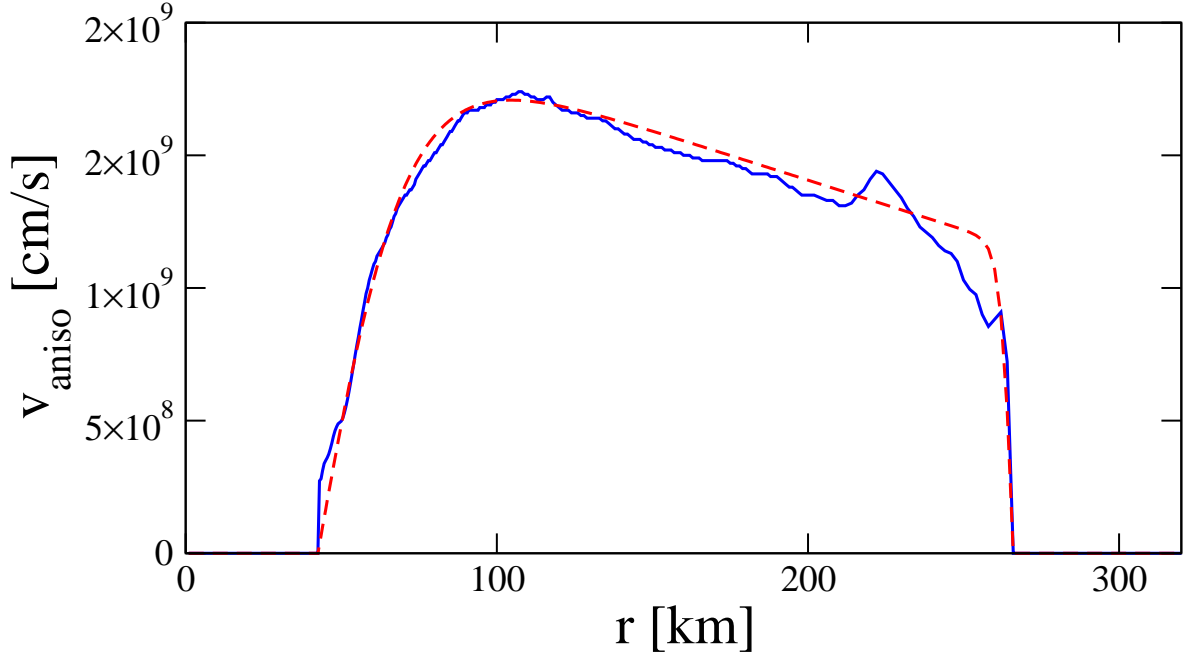}
\caption{The anisotropic velocity as a function of the radius $r$. The data from the Couch \& O'Connor simulation \cite{2014ApJ...785..123C} at $t=170\;{\rm ms}$ postbounce is shown as the solid blue line. The fit, given by equation (\ref{eq:vanisofit}), is the dashed red line. \label{fig:vaniso170ms}}
\end{figure}
For our calculations, we adopt an envelope function $\Omega(r)$ based upon the density-weighted anisotropic velocity $v_{aniso}$ of material within the shock as a function of the radius $r$. The anisotropic velocity is defined to be \cite{2012ApJ...749...98T} 
\begin{equation}
 v_{aniso} = \sqrt{\frac{\langle\rho\left[(v_r - \langle v_r\rangle)^{2} + v_{\theta}^2 + v_{\phi}^2\right]\rangle}{\langle\rho\rangle} },
 \label{eq:vaniso}
\end{equation}
where $v_r$, $v_{\theta}$ and $v_{\phi}$ are the radial, azimuthal and polar velocities of the fluid at $r$, and $\rho$ is the mass density at $r$. The averages indicated by the angle brackets are over the angular coordinates. Since the Fischer \emph{et al.} simulations are one-dimensional (and thus cannot possess anisotropic velocities), we compute $v_{aniso}$ from 8 snapshots of the 3D simulations found in Couch and O'Connor \cite{2014ApJ...785..123C}. The calculated $v_{aniso}$ at $t=170\;{\rm ms}$ from the simulation is shown as the blue solid line in figure (\ref{fig:vaniso170ms}), where we see $v_{aniso}$ is non-zero in the region between the proto-neutron star surface $R_{NS} = 43\;{\rm km}$ and the shock radius $R_S=264 \;{\rm km}$, reaching a maximum around $r \sim 100\;{\rm km}$ which is close to the gain radius at $R_G = 65\;{\rm km}$ for this snapshot. We find the anisotropic velocity at all the snapshots we examined can be fit with a function of the form
\begin{equation}
v_{aniso}(r) \propto \tanh\left(\frac{r-R_{NS}}{\Lambda_{NS}}\right)\,\tanh\left(\frac{R_S-r}{\Lambda_S} \right) \left( 1-m(r-R_{NS}) \right), \label{eq:vanisofit}
\end{equation}
where $\Lambda_{NS}$ and $\Lambda_S$ are two damping scales. The fit to $v_{aniso}$ at $t=170\;{\rm ms}$ from the Couch and O'Connor simulation is shown as the red dashed line in figure (\ref{fig:vaniso170ms}). From the fits to $v_{aniso}$ at the eight snapshot times we analysed, we find $\Lambda_{NS}$ and $\Lambda_{S}$ approach asymptotes of $\Lambda_{NS}=16\;{\rm km}$ and $\Lambda_S=7\;{\rm km}$ for postbounce times beyond $t=250\;{\rm ms}$. The slope $m$, the neutron star radius $R_{NS}$, and the gain radius $R_G$ can be fit as a function of the shock radius $R_S$ and are given by:
\begin{eqnarray}
 R_{NS} & = & 6.50\times 10^{8} \left(\frac{1\;{\rm cm}}{R_S}\right)^{0.29}\;{\rm cm}, \\ 
 R_G & = & 9.40\times 10^{8} \left(\frac{1\;{\rm cm}}{R_S}\right)^{0.28}\;{\rm cm}, \\  
 m & = & max(0, 3.26 \times 10^{-8} - 2.97\times 10^{-16}\;{\rm cm^{-1}}\,R_S)
\end{eqnarray}
This form of $m$ is used to avoid the possibility that the gradient $m$ becomes negative for large shock radii. 

\begin{figure}[t!] 
\includegraphics[width=0.7\linewidth]{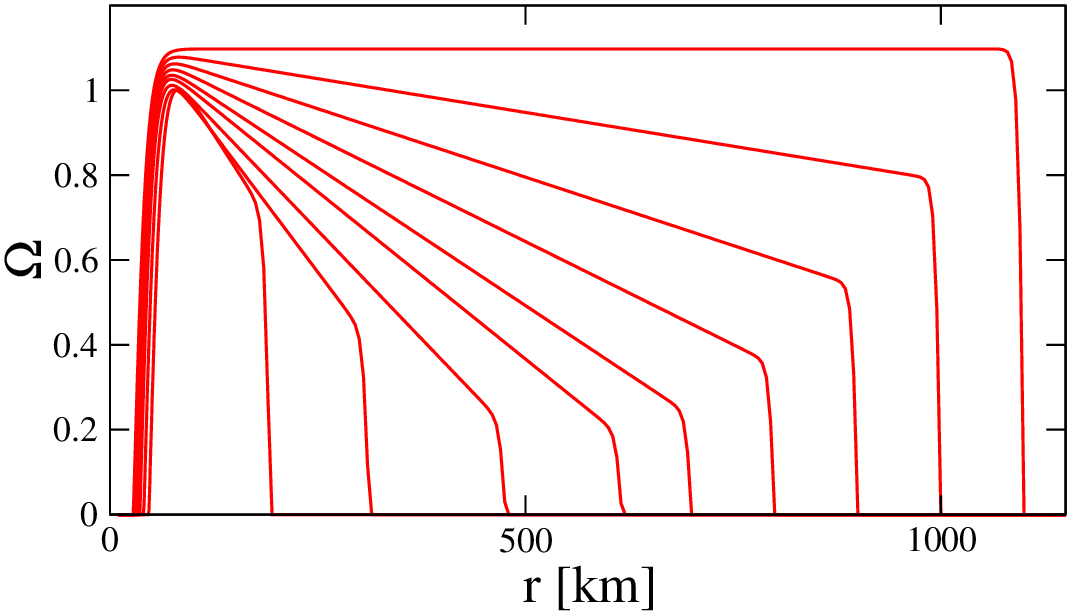}
\caption{The envelope function $\Omega(r;R_S)$ given by equation (\ref{eq:Omega}) as a function of shock radius. At early times/small shock radii the turbulence is largest in the region around the gain radius; at later times/larger shock radii the turbulence becomes homogeneous.
\label{fig:Omega}}
\end{figure}
\begin{figure}[t!] 
\includegraphics[width=0.97\linewidth]{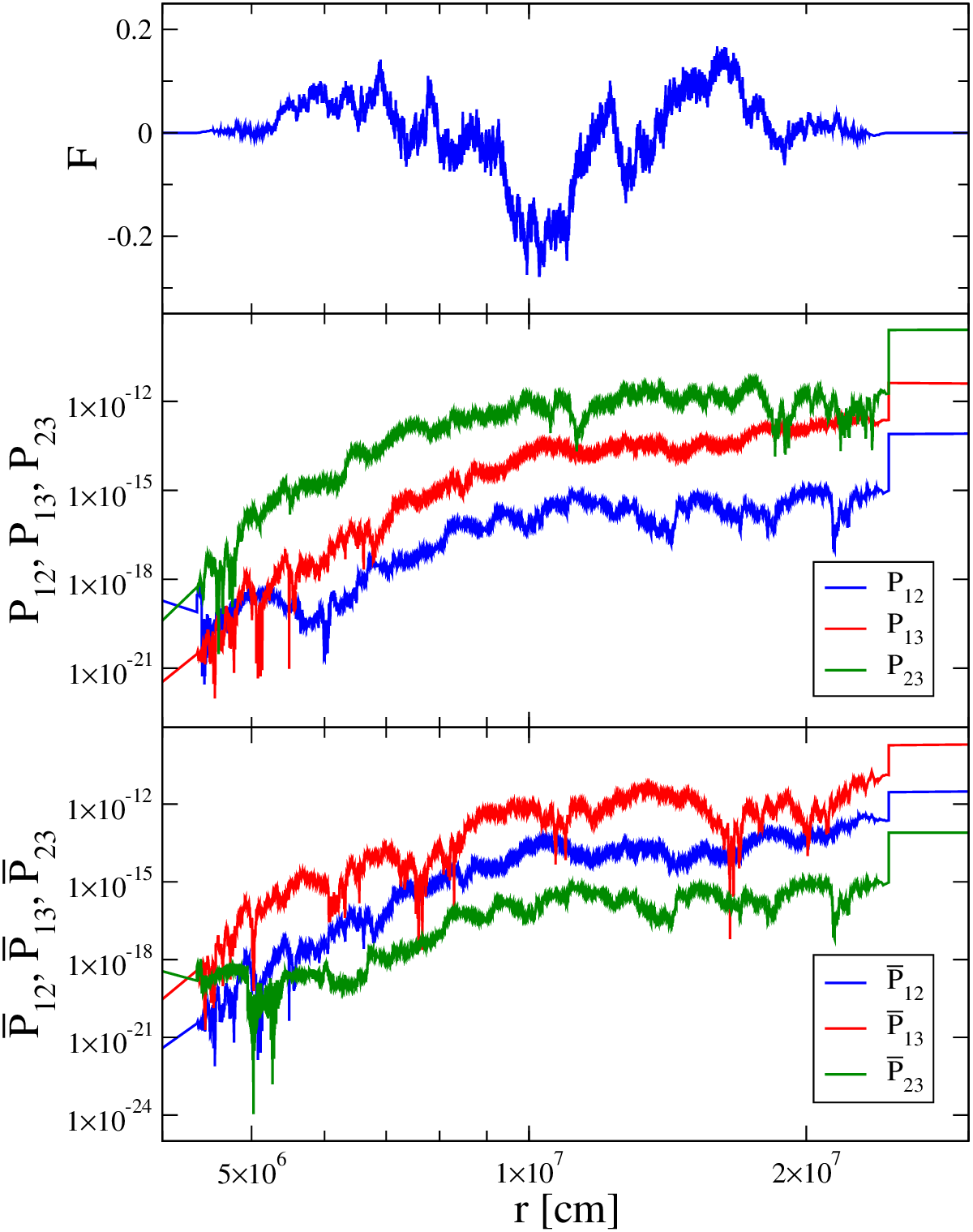}
\caption{The random field $F$ (top panel), the neutrino transition probabilities $P_{12}$, $P_{13}$ and $P_{23}$ (middle panel), and the antineutrino transition probabilities $\bar{P}_{12}$, $\bar{P}_{13}$ and $\bar{P}_{23}$ (bottom panel) as a function of distance $r$. For the lower two panels the colour coding is: $P_{12}$ and $\bar{P}_{12}$ are blue,  $P_{13}$ and $\bar{P}_{13}$ are red, $P_{23}$ and $\bar{P}_{23}$ are green. The mass density profile is the $t=0.3\;{\rm s}$ snapshot shown in figure (\ref{fig:rhoYe}).
\label{fig:Pvsr465}}
\end{figure}
The envelope function $\Omega(r;R_S)$ we adopt for the calculations is taken to be the anisotropic velocity normalized to the gain radius $R_G$:
\begin{equation}
 \Omega(r;R_S) = \frac{v_{aniso}(r)}{v_{aniso}(R_G)}
 \label{eq:Omega}
\end{equation}
A plot of $\Omega(r;R_S)$ at various shock radii is shown in figure (\ref{fig:Omega}). As the shock moves outwards, the gain radius and the neutrinosphere radius slowly contract and the gradient decreases. The net effect is that as a function of shock radius, the turbulence becomes more homogeneous as $R_S$ increases.

Finally, we mention that the algorithm used to generate the random wavenumbers and amplitudes is the same as described in Kneller \& Mauney \cite{2013PhRvD..88b5004K}. In order to cover all spatial scales down to the oscillation wavelength close to the proto neutron star, we generate wavenumbers over ten orders of magnitude: from $q_{cut}$ to $10^{10}\,q_{cut}$ with fifty wavenumbers per decade. 


\subsection{Numerical Results}
\begin{figure}[t!] 
\includegraphics[width=\linewidth]{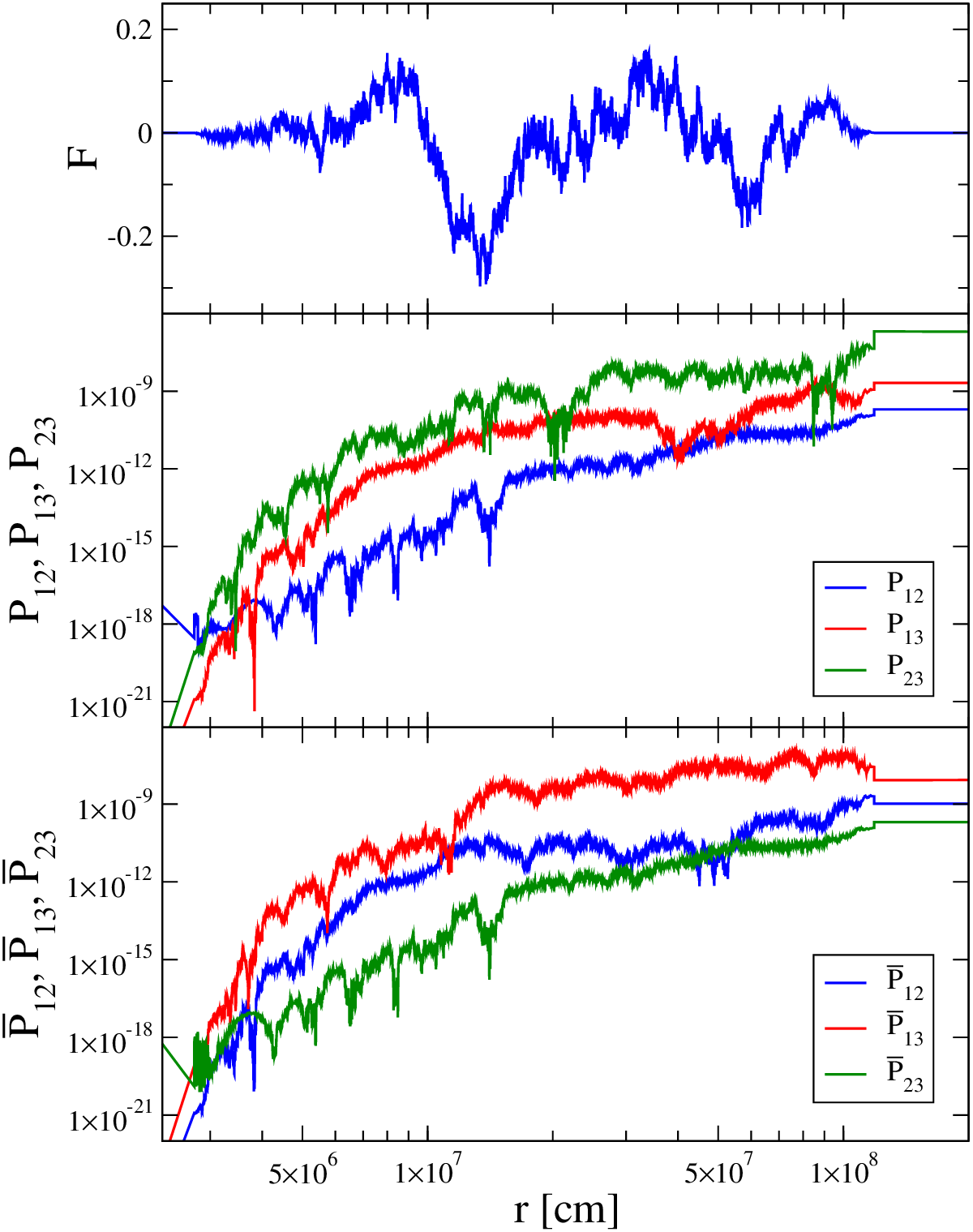}
\caption{The same as in figure (\ref{fig:Pvsr465}) but for the mass density profile at $t=0.5\;{\rm s}$ snapshot shown in figure (\ref{fig:rhoYe}).
\label{fig:Pvsr492}}
\end{figure}
\begin{figure}[t!] 
\includegraphics[width=\linewidth]{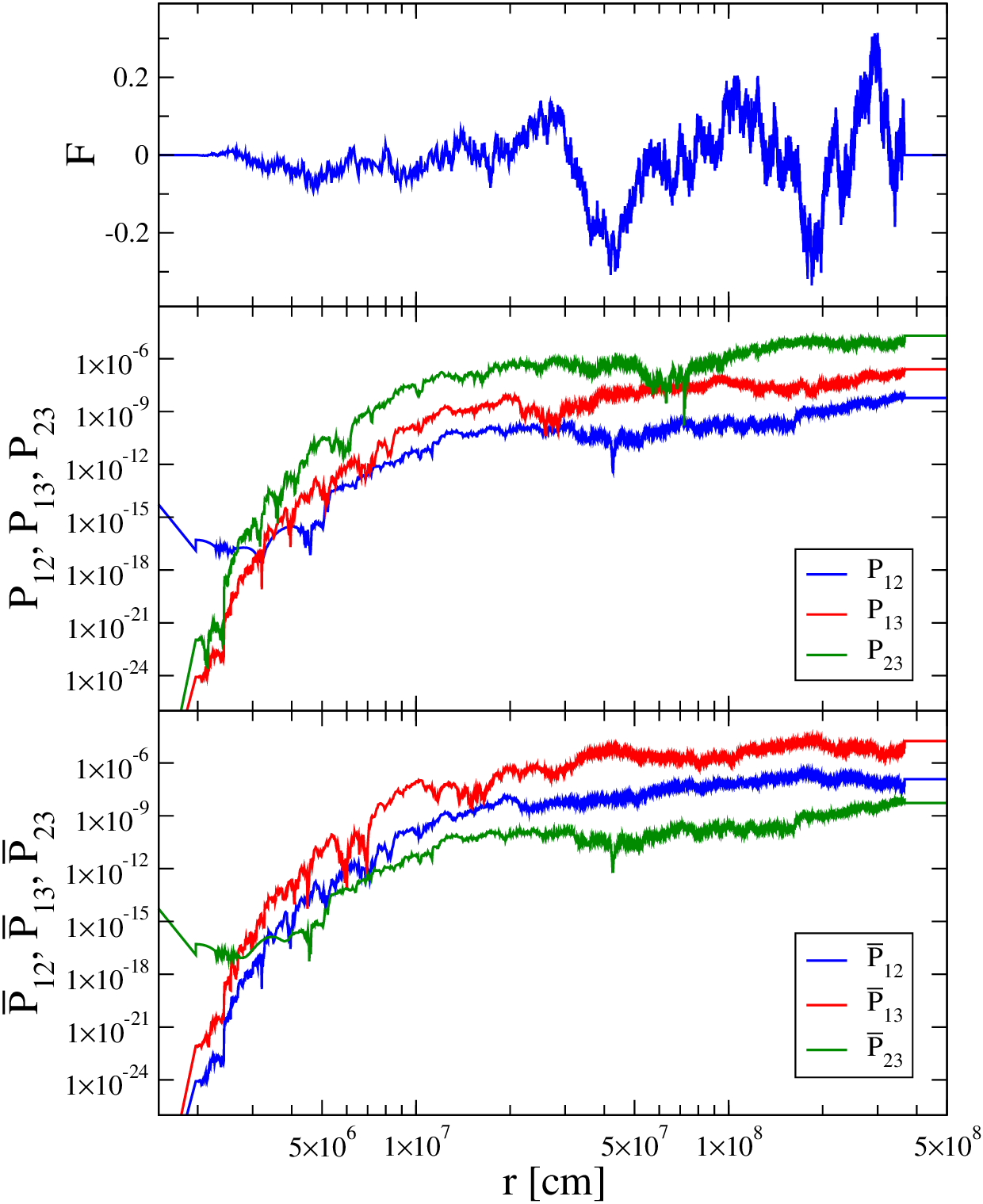}
\caption{The same as in figure (\ref{fig:Pvsr465}) but for the mass density profile at $t=0.7\;{\rm s}$ snapshot shown in figure (\ref{fig:rhoYe}). 
\label{fig:Pvsr509}}
\end{figure}

Armed with all the pieces that go into the calculations described above, we now undertake a set of numerical calculations for the transition probabilities of both a neutrino and antineutrino as a function of distance through the $t=0.3\;{\rm s}$, $t=0.5\;{\rm s}$ and $t=0.7\;{\rm s}$ profiles shown in figure (\ref{fig:rhoYe}). The transition probabilities are defined to be the probability that a neutrino with initial generic state $\nu_j$ is found in generic state $\nu_i$ after traveling a distance $r$, i.e.\ $P(\nu_j \rightarrow \nu_i)$. If the two states are in the same basis and the evolution matrix $S$ in that basis is known, then $P(\nu_j \rightarrow \nu_i) = P_{ij} = |S_{ij}|^2$ where $S_{ij}$ is the $ij$'th element of $S$. We shall report transition probabilities in the instantaneous eigenstate basis -- that is, the `matter' basis \cite{2012JPhG...39c5201G} -- because these probabilities are approximately constant in the absence of turbulence except at discontinuities. Antineutrino transition probabilities will be denoted as $P(\bar{\nu}_j \rightarrow \bar{\nu}_i) = \bar{P}_{ij}$. We consider a neutrino energy of $E=10\;{\rm MeV}$ and use the following mixing parameters: $m_2^2 - m_1^2 = 7.5\times 10^{-5}\;{\rm eV^2}$, $m_3^2 - m_2^2 = 2.32\times 10^{-3}\;{\rm eV^2}$, $\theta_{12} = 33.9^{\circ}$, $\theta_{13} = 9^{\circ}$, $\delta_{13}=0$, $\theta_{23}=45^{\circ}$ \cite{Olive:2016xmw}. The phase $\delta_{13}$ does not affect neutrino oscillations if the radiative corrections \cite{2008PhRvD..77f5024E} to the matter Hamiltonian are ignored \cite{2008PhLB..662..396B,2008PhRvD..78h3007G,2009PhRvD..80e3002K}. With these mixing parameters and energy one finds two MSW resonances occur: one between states $\nu_1$ and $\nu_2$ at a mass density of $\rho_{MSW,12} \approx 38\;{\rm g/cm^3}$ and another between $\nu_2$ and $\nu_3$ at a mass density of $\rho_{MSW,23} \approx 3000\;{\rm g/cm^3}$. These densities are well below those behind the shock during the accretion phase seen in figure (\ref{fig:rhoYe}). Although we use the normal neutrino mass mass ordering here, we have also undertaken inverted mass ordering calculations and found no qualitative differences. The turbulence amplitude is set to $C_{\star} = 0.1$ and the power spectrum index to the Kolmogorov value $\alpha = 5/3$. 

The results are shown in figures (\ref{fig:Pvsr465}), (\ref{fig:Pvsr492}) and (\ref{fig:Pvsr509}). 
For all three mass density profiles used, we see the turbulence effects for both neutrinos and antineutrinos are comparable in scale. As the shockwave moves outward, the effects due to the turbulence grow in size and phenomenologically we find the curves for $P_{ij}(r)$ are similar to the curves for $1/V^{C}(r)$. But we also observe that in none of the figures do the transition probabilities reach order unity. Thus we conclude that for these profiles, this neutrino energy and mixing parameters, and these properties of the turbulence, no significant effect from turbulence occurs. In other words, turbulence would have had no effect upon the neutrino heating in the simulation, even if the simulation had a spatial resolution as small as $\sim 10\;{\rm \mu m}$. The obvious question is why not, and the next is what would it take for turbulence to have an effect? To answer those questions we use the analytical models for turbulence which we now describe.


\section{Stimulated Transitions} \label{sec:thescales}

As shown by Kneller \& Kabadi \cite{2015PhRvD..92a3009K}, turbulence affects the neutrino flavour evolution via two different paths: Stimulated Transition, and Distorted Phase Effects. We begin with the Stimulated Transition model and describe Distorted Phase Effects in the next section.

The Stimulated Transition model treats the neutrino like a polar molecule with eigenstates given by a non-turbulent Hamiltonian, and the turbulence as an external potential that drives transitions between those eigenstates. This separation of the Hamiltonian into turbulence-free and turbulence-filled components aligns closely with the method by which turbulence was modeled in the numerical calculations. We separate the matter potential in the flavour basis, $H^{(f)}_M$, into a smooth, turbulence free, component, $\breve{H}_{M}^{(f)}$ that depends upon the turbulence-free potential $\breve{V}^{\,C}$, and the turbulence filled perturbation $\delta H^{(f)}_M$. The eigenvalues of the unperturbed Hamiltonian, given by $\breve{H} = H_V + \breve{H}_{M}$, are $k_i$, and we can also introduce an unperturbed mixing matrix $U$ which diagonalizes the unperturbed Hamiltonian $\breve{H}$. 
\begin{figure*}[t!] 
\includegraphics[width=0.49\linewidth]{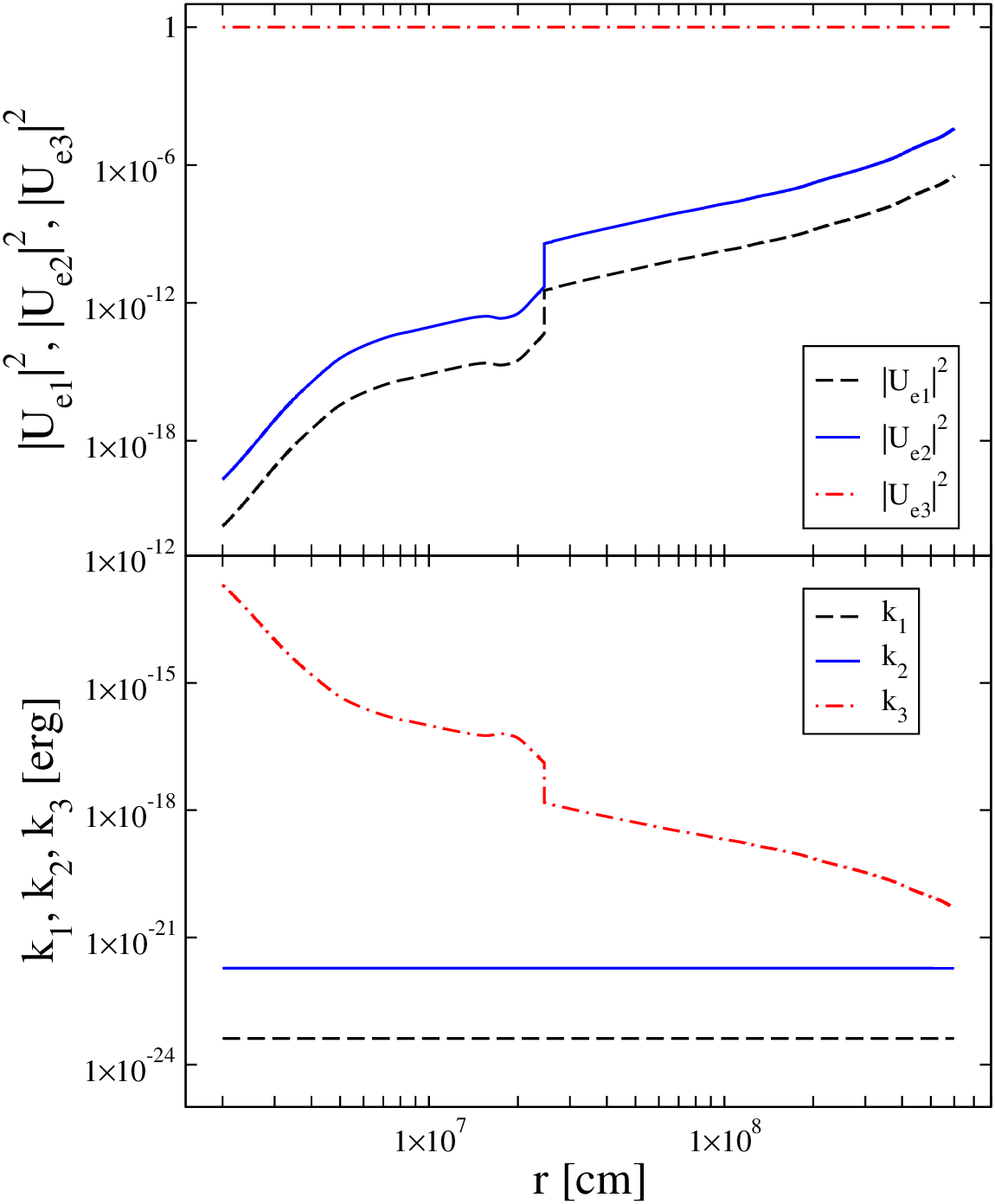}
\includegraphics[width=0.49\linewidth]{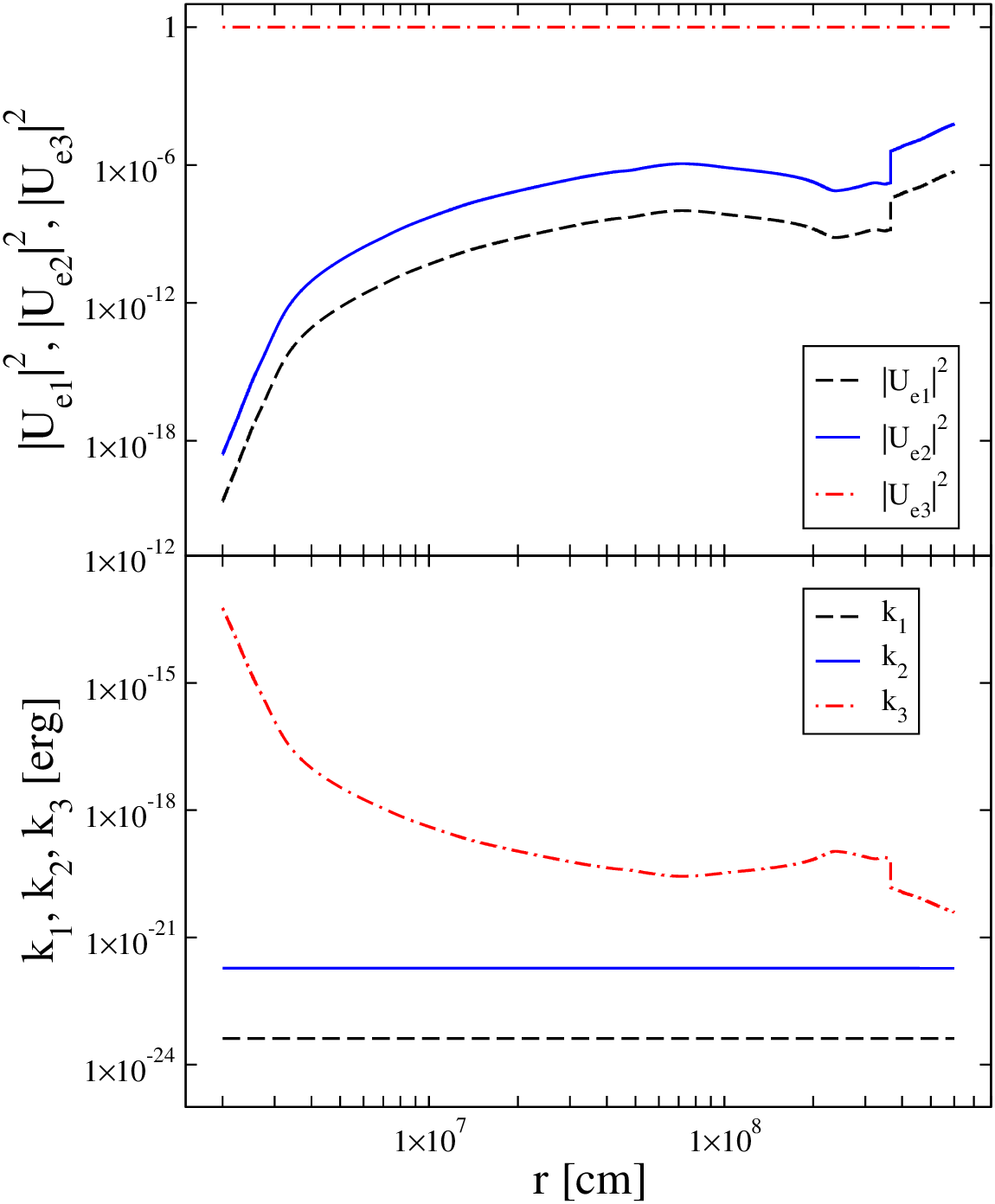}
\caption{The square magnitude of the unperturbed mixing matrix elements $|U_{e1}|^2$, $|U_{e2}|^2$, $|U_{e3}|^2$ (top panels) and the eigenvalues of the unperturbed Hamiltonian $k_1$, $k_2$ and $k_3$ (bottom panels) for the $t=0.3\;{\rm s}$ (left) and $t=0.7\;{\rm s}$ (right) profiles shown in figure (\ref{fig:rhoYe}). $|U_{e1}|^2$ and $k_1$ are denoted by the black dashed line, $|U_{e2}|^2$ and $k_2$ by the blue solid lines, and  $|U_{e3}|^2$ and $k_3$ by the red dot-dashed lines.\label{fig:ks,Us,3flavour}}
\end{figure*}
Figure (\ref{fig:ks,Us,3flavour}) shows the square magnitude of the unperturbed mixing matrix elements $|U_{e1}|^2$, $|U_{e2}|^2$, $|U_{e3}|^2$ and the eigenvalues of the unperturbed Hamiltonian $k_1$, $k_2$ and $k_3$ as a function of distance $r$ through the $t=0.3\;{\rm s}$ and $t=0.7\;{\rm s}$ snapshots from figure (\ref{fig:rhoYe}). One observes how at both times $|U_{e3}|^2$ is constant while $|U_{e1}|^2$ and $|U_{e2}|^2$ are proportional to $1/\breve{V}^{\,C}$; similarly, $k_3$ is proportional to $\breve{V}^{\,C}$, and $k_1$ and $k_2$ are constant. These behaviors are expected from the formulae given in Kneller \& McLaughlin \cite{2009PhRvD..80e3002K} for the eigenvalues and matter mixing angles in the limit of high mass density. For the antineutrinos we find the differences between the unperturbed eigenvalues, denoted by $\bar{k}_{i}$, at these densities are such that $\bar{k}_{2}-\bar{k}_{1} = k_{3}-k_{2}$, $\bar{k}_{3}-\bar{k}_{1} = k_{3}-k_{1}$ and $\bar{k}_{3}-\bar{k}_{2} = k_{2}-k_{1}$, while the elements of the antineutrino unperturbed mixing matrix, which we denote by $\bar{U}$, are such that $|\bar{U}_{e1}|^2 = |U_{e3}|^2$, $|\bar{U}_{e2}|^2 = |U_{e2}|^2$, and $|\bar{U}_{e3}|^2 = |U_{e1}|^2$.   
For an inverted mass ordering the eigenvalues and mixing matrix elements in this region look essentially the same with reassignments $k_1\rightarrow k_3$, $k_2\rightarrow k_1$, $k_3\rightarrow k_2$, $|U_{e1}|^2\rightarrow |U_{e3}|^2$, $|U_{e2}|^2\rightarrow |U_{e1}|^2$, $|U_{e3}|^2\rightarrow |U_{e2}|^2$. 

What the figures also make clear is there are no MSW resonances: at no location for $r < R_{S}$ do we find the mixing matrix elements changing from $|U_{ei}|^2 \sim 0$ to $|U_{ei}|^2 \sim 1$ (or vice versa) or the difference between eigenvalues $|k_i - k_j|$ passing through a minimum, both of which occur at MSW resonances. This absence of MSW resonances indicates the evolution of neutrinos and antineutrinos due to the unperturbed Hamiltonian should be very close to adiabatic. 

Now that we know the evolution of the eigenstates of the unperturbed Hamiltonian, we analyze the effect of the turbulence upon them using the Stimulated Transition model. The original description of the model \cite{Patton:2013dba} assumed the unperturbed Hamiltonian was a constant. In these circumstances one finds five different lengthscales emerge: the cutoff scale $\lambda_{cut}$ which is the longest wavelength of the turbulence; the turbulence dissipation scale $\lambda_{diss}$ which is the shortest wavelength of the turbulence, the splitting scale $\lambda_{split,ij}$ which is the wavelength corresponding to the energy difference between pairs of unperturbed eigenvalues $i$ and $j$; $\lambda_{trans,ij}$ which is the lengthscale associated with the transitions between the states $i$ and $j$ when the neutrino is on resonance; and $\lambda_{ampl,ij}$ which is the wavelength of the turbulence which suppresses transitions between eigenstates $i$ and $j$. In order for turbulence to have a sizable effect, two different criteria must be satisfied between the five lengthscales. Later Patton, Kneller \& McLaughlin \cite{Patton:2014lza} showed how to apply the model when the unperturbed Hamiltonian has a spatial dependence which introduces a sixth lengthscale, the potential scale height $h^{\,C}$. Now, in order to have a sizable effect, a third criteria must be satisfied. Before we describe the three criteria, we consider the six lengthscales in more detail. 

\subsection{$\lambda_{cut}$}

The first scale we need to consider is $\lambda_{cut}$, the longest wavelength of the turbulence. In what follows we use $\lambda_{cut} = 2 R_S$, i.e.\ a wavelength which is the diameter of the region behind the shock. This is consistent with the definition of $q_{cut}$ used in the numerical calculations shown in section \S\ref{sec:neutrinopropagation}. Using the profiles shown in figure (\ref{fig:rhoYe}) we find $\lambda_{cut}$ is on the order of $500\;{\rm km}$ to $5,000\;{\rm km}$. For 3D turbulence, $\lambda_{cut}$ is the distance scale at which turbulence is generated, but for 2D turbulence this is not the case due to the possibility of the inverse cascade. In three dimensions, $\lambda_{cut}$ is usually not the largest amplitude turbulent mode, which is typically found to be on a scale an order of magnitude smaller than $\lambda_{cut}$ \cite{2015ApJ...808...70A}. 

\subsection{$\lambda_{diss}$}

The second scale is the turbulence dissipation scale, or the shortest wavelength of the turbulence. This scale can be estimated using equations (20) and (31) from Abdikamalov \emph{et al.} \cite{2015ApJ...808...70A} after adopting conservative values of $\rho = 10^9\;{\rm g/cm^3}$ for the mass density, $v = 10^8\;{\rm cm/s}$ for fluid velocity, $T = 1\;{\rm MeV}$ for the temperature, and a characteristic size of the turbulent region of $R_S$. We find $\lambda_{diss}$ to be $\lambda_{diss} \approx 10^{-5}\;{\rm cm}$. Other effects may quench the turbulence power spectrum at longer wavelengths than our estimate so we regard this estimate of $\lambda_{diss} \approx 10^{-5}\;{\rm cm}$ as being a lower limit. The ratio of $\lambda_{cut}$ to $\lambda_{diss}$ is of order $10^{13}$, a huge dynamic range that would be difficult to capture in a hydrodynamical simulation. 

\subsection{$\lambda_{split,ij}$}

The next scale we need to define is $\lambda_{split,ij}$, which are the wavelengths corresponding to the differences between given pairs of eigenstates of the unperturbed Hamiltonian. Just as with atoms or molecules interacting with photons, transitions between neutrino states will occur with greater amplitude when there are Fourier modes in the turbulence which match these differences between eigenvalues. When this occurs, one can describe the states as being resonant -- not to be confused with MSW resonances. In principle it is possible to combine two or more turbulence modes to achieve resonance \cite{Patton:2014lza}, but we shall not consider that possibility here because such resonances tend to be very narrow. For a three-flavor neutrino there are three eigenvalue differences, so there are three scales for $\lambda_{split,ij}$. For a given pair of states $i$ and $j$, the splitting scale is given by 
\begin{equation}
\lambda_{split,ij} = \frac{2\pi}{|k_{i} - k_{j}|}.
\label{eq:lambdasplit}
\end{equation}
From figure (\ref{fig:ks,Us,3flavour}), which shows the unperturbed eigenvalues at two different snapshots, we see the difference $|k_{3} - k_{2}| \approx |k_{3} - k_{1}| \approx \breve{V}^{\,C}$ while $|k_{2} - k_{1}|$ is a constant. The differences for the antineutrinos are not shown but are found to be $|\bar{k}_{3} - \bar{k}_{2}| \approx |k_{2} - k_{1}|$, $|\bar{k}_{3} - \bar{k}_{1}| \approx |\bar{k}_{2} - \bar{k}_{1}| \approx \breve{V}^{\,C}$. These splittings are for a normal mass ordering; for an inverse mass ordering the splittings are approximately the same, with the substitutions $3\rightarrow 2$, $2\rightarrow 1$, $1\rightarrow 3$ for both neutrinos and antineutrinos. So, for example, in an inverted mass ordering $|k_{2} - k_{1}|$ is approximately the same as the splitting $|k_{3} - k_{2}|$ in a normal mass ordering. 

\subsection{$\lambda_{trans,ij}$}

The next scale we introduce is the lengthscale over which the neutrino makes transitions between the eigenstates of the unperturbed Hamiltonian. This scale is found by solving for the evolution of a quantum system subject to a Fourier-decomposed perturbation. The procedure for deriving this scale is given in detail in Patton, Kneller \& McLaughlin \cite{Patton:2013dba} and Yang, Kneller \& Perkins \cite{2015arXiv151001998Y} so we include only a simple sketch of the calculation. In the eigenbasis of the unperturbed Hamiltonian, the evolution matrix evolves according to the Hamiltonian 
\begin{equation}
H^{(u)} = K - \rmi U^{\dagger}\,\frac{dU}{dr} + U^{\dagger}\delta H^{(f)} U
\end{equation} 
where $K$ is the diagonal matrix of eigenvalues of the unperturbed Hamiltonian, $K= diag(k_1,k_2,k_3)$ and $U$ the previously introduced unperturbed mixing matrix. We write the evolution matrix in the unperturbed eigenbasis as the product $S = \breve{S}\,A$ where $\breve{S}$ is defined to be the solution of 
\begin{equation}
\label{eq:dbreveSdt}
\rmi \frac{d\breve{S}}{dr} = \left[ K - \rmi \breve{U}^{\dagger}\,\frac{d\breve{U}}{dr} \right] \,\breve{S}.
\end{equation}
For a constant unperturbed Hamiltonian the solution of this equation is $\breve{S}=\exp\left(-\rmi K r\right)$. 
The evolution equation for $A$ is thus
\begin{equation}
\rmi \frac{dA}{dr} = \breve{S}^{\dagger}\,U^{\dagger}\delta H^{(f)}U\,\breve{S} \,A. \label{dAdt}.
\end{equation}
The term $U^{\dagger}\delta H^{(f)}U$ which appears in this equation in general possesses both
diagonal and off-diagonal elements. The diagonal elements can be removed by writing the matrix $A$ as $A=W\,B$ where $W=\exp(-\rmi\Xi)$ and $\Xi$ is a diagonal matrix $\Xi=\rm{diag}(\xi_{1},\xi_{2},\xi_{3})$. Using this expression for $A$ in equation (\ref{dAdt}) gives a differential equation for $B$
\begin{equation}\label{dBdt}
\rmi \frac{dB}{dr} = W^{\dagger}\left[\breve{S}^{\dagger}U^{\dagger}\delta H^{(f)}U\,\breve{S} -\frac{d\,\Xi}{dr}\right]\,W\,B \equiv H^{(B)}\,B
\end{equation}
and $\Xi$ is chosen so that $d\,\Xi/dr$ removes the diagonal elements of $\breve{S}^{\dagger}U^{\dagger}\delta H^{(f)}U\,\breve{S}$.
If the perturbations in the potential are decomposed into Fourier series of the form
\begin{equation}
\delta V^{\,C} = \breve{V}^{\,C}\,\sum_a^{N_q}\,G_{a}\,\sin\left(q_a r + \eta_a \right)
\end{equation}
where $q_a$ is the wavenumber, $\eta_a$ is a phase, and $G_a$ is the amplitude, then 
we can find an analytic solution for $\Xi$:
\begin{equation}
\xi_i = \breve{V}^{\,C} |U_{ei}|^2\,\sum_a^{N_q} \frac{G_a}{q_a}\left[ \cos\eta_a - \cos\left(q_a r+\eta_a\right) \right]
\end{equation}
when $\breve{V}^{\,C}$ is constant.
In order to solve equation (\ref{dBdt}) for $B$ we have to make an approximation known as the Rotating Wave Approximation. This approximation amounts to finding a set of $N_q$ integers, one for each Fourier mode $a$ and each off-diagonal element $ij$ of the perturbing Hamiltonian. We call these integers $n_{a;ij}$, and the complete sets $\left\{ n_{ij}\right\}$ are the RWA integers for off-diagonal element $ij$. The RWA integers do not have to be the same for every off-diagonal element $ij$ but the sets are not all independent: the integer for Fourier mode $a$ in the set for element $ij$ is related to the integer for mode $a$ in the sets for elements $ik$ and $kj$ by $n_{a;ij} = n_{a;ik} + n_{a;kj}$. For three flavors this means the set for one off-diagonal element is given in terms of the sets for the other two. 

After using the Rotating Wave Approximation we find the Hamiltonian for $B$ is of the form
\begin{equation} \label{eq:HBRWA}
\fl
H^{(B)} = \left( \begin{array}{ccc}
0 & \kappa_{12}\,e^{\rmi \left[ \delta k_{12} + \sum\limits_a n_{a;12}\,q_a \right]\,r} &  \kappa_{13}\,e^{\rmi \left[ \delta k_{13} + \sum\limits_a n_{a;13}\,q_a \right]\,r}\\
\kappa^{\star}_{12}\,e^{-\rmi \left[ \delta k_{12} + \sum\limits_a n_{a;12}\,q_a \right]\,r} & 0 & \kappa_{23}\,e^{\rmi \left[ \delta k_{23} + \sum\limits_a n_{a;23}\,q_a \right]\,r} \\
\kappa^{\star}_{13}\,e^{-\rmi \left[ \delta k_{13} + \sum\limits_a n_{a;13}\,q_a \right]\,r} & \kappa^{\star}_{23}\,e^{-\rmi \left[ \delta k_{23} + \sum\limits_a n_{a;23}\,q_a \right]\,r} & 0
\end{array} \right)
\end{equation}
where $\delta k_{ij} = k_i -k_j$.
In this equation the quantities $\kappa_{ij}$ are 
\begin{equation}
\kappa_{ij} = \frac{U_{ei}^{\star}\,U_{ej}}{\left(|U_{ei}|^2- |U_{ej}|^2\right)}  \sum_{a=1}^{N_q}\left\{ n_{a;ij}\,q_{a} \right\} \prod_{b=1}^{N_q} J_{n_{b;ij}}\left(z_{a;ij}\right)  
\label{eq:kappaij}
\end{equation} 
with $J_n$ Bessel functions and the quantities $z_{a;ij}$ given by
\begin{equation}
 z_{a;ij} =  \frac{G_{a}\,\breve{V}^{\,C}}{q_a}\,\left(|U_{ei}|^2- |U_{ej}|^2\right)
\end{equation}
Although equation (\ref{eq:HBRWA}) looks complicated, it actually has a general solution. To evaluate it we need the amplitude and wavenumber of every mode in the turbulence and the set of RWA integers for each element in order to compute the quantities $z_{a;ij}$ and thus $\kappa_{ij}$. The solution is not very intuitive, so let us make a pair of reasonable-sounding approximations that allow us to estimate the scale over which transitions between states are made. 

First, let us focus upon just two states at a time and ignore the third. In this limit we find the transition probability between the two states varies as $P_{ij} = \frac{|\kappa_{ij}|^2}{Q_{ij}^2}\,\sin^{2}\left( Q_{ij} r\right)$ where $Q_{ij}^2 = p_{ij}^2 + |\kappa_{ij}|^2$ and $2\,p_{ij} =  \delta k_{ij} + \sum\limits_a n_{a;ij}\,q_a$ is the detuning frequency. This sinusoidal transition probability defines the wavelength $\lambda_{Q,ij} = 2\pi/Q_{ij}$. Even with a two-state approximation, this wavelength still depends upon a lot of different information. Our next approximation is to assume that for every pair of states $i$ and $j$ with eigenvalue difference $\delta k_{ij}$ there is a turbulence mode with an exactly matching wavenumber. In this case we have a resonance condition and the detuning frequency is exactly zero. Under these circumstances 
\begin{equation}
\lambda_{Q,ij} = \frac{2\pi}{|\kappa_{ij}|} 
\end{equation}
It was shown by Patton, Kneller \& McLaughlin \cite{Patton:2014lza} that large effects occur only close to the resonance ($p < |\kappa|$), justifying this approximation. Since there is a turbulence mode which exactly matches $\delta k_{ij}$, the RWA integers for mixing between this pair of states will be all zeros except for that one Fourier mode where $q_a = |\delta k_{ij}|$ which will have $|n_a|=1$; the sign is determined by ordering of the eigenvalues. The next approximation we make is the small amplitude approximation, that $z_{a;ij} \ll 1$ for all modes $a$. In this limit $J_0(z) \approx 1$ and $J_1(z) \approx z/2$. 

Putting these approximations together and denoting the amplitude of the Fourier mode on resonance by $G_{\star,ij}$, we find the combination $G_{\star,ij}\,\lambda_{Q,ij} > \lambda_{trans,ij}$ on resonance where $\lambda_{trans,ij}$ is a scale we define to be
\begin{equation}
\lambda_{trans,ij} = \frac{4\pi}{ \breve{V}^{\,C}\,|U_{ei}^{\star}\,U_{ej}| }.
\label{eq:lambdatrans}
\end{equation}
Figure (\ref{fig:ks,Us,3flavour}) allows us to get a feel for $\lambda_{trans,ij}$ as a function of distance $r$ for different pairs of states. The figure shows that both $|U_{e1}|$ and $|U_{e2}|$ are proportional to $1/\breve{V}^{\,C}$ whereas $|U_{e3}|$ is very close to unity. Thus both products $\breve{V}^{\,C}\,|U_{e1}^{\star}\,U_{e3}|$ and $\breve{V}^{\,C}\,|U_{e1}^{\star}\,U_{e3}|$ are constant while $\breve{V}^{\,C}\,|U_{e1}^{\star}\,U_{e2}|$ is proportional to $1/\breve{V}^{\,C}$ so will diverge as the mass density becomes large close to the proto-neutron star. The lack of dependence of $\lambda_{trans,ij}$ on $\breve{V}^{\,C}$ in the $1-3$ and $2-3$ mixing channels is important to note: the lack of turbulence effects in figures (\ref{fig:Pvsr465}), (\ref{fig:Pvsr492}) and (\ref{fig:Pvsr509}) is not because the transition wavelength diverges in dense matter due to alignment of flavour and the instantaneous eigenstates. If the amplitude of the fluctuations in the potential are proportional to the potential, as we have assumed, the transition wavelength is constant in these channels. 

If $\lambda_{trans,ij}$ in some channels is constant, the relation
\begin{equation}
\lambda_{Q,ij} \geq \frac{\lambda_{trans,ij}}{G_{\star,ij}} \label{eq:lambdatrans3}
\end{equation}
shows that the wavelength $\lambda_{Q,ij}$ is inversely proportional to the amplitude of the resonant Fourier mode. For a given realization of turbulence, the amplitudes of the Fourier modes are random, but we expect $G_{\star,ij} \approx C_{\star}\,\Omega(r)\,E(\delta k_{ij})$. For an inverse power law turbulence power spectrum, 
\begin{equation}
G_{\star,ij} \propto \left(\frac{q_{cut}}{\delta k_{ij}}\right)^{\alpha}.
\end{equation}
We find that the smaller the difference between the eigenvalues, the larger the amplitude of the resonant Fourier mode, and thus the smaller the transition wavelength $\lambda_{Q,ij}$. 

\subsection{$\lambda_{ampl,ij}$}

The small amplitude approximation used to derive $\lambda_{trans,ij}$ does not always hold. Even if the RWA integer for a particular mode is zero, that mode can still influence the wavelength $\lambda_{Q,ij}$. This dependence upon all Fourier modes appears via the Bessel functions that appear in equation (\ref{eq:kappaij}). If the mode $a$ does not contribute to the resonance condition, then $\kappa_{ij} \propto J_{0}\left(z_{a;ij}\right)$. When the argument $z_{a;ij}$ of the Bessel function is greater than the first root of $J_0$, i.e.\ $z_{a;ij} \geq 2.4048...$, then $J_{0}$ evaluates to a small number, $\kappa_{ij}$ will be suppressed, and the actual wavelength $\lambda_{Q,ij}$ will be \emph{much} greater than the lower limit $\lambda_{trans,ij}/G_{\star,ij}$ derived in the previous section. From the definition of $z$ we find this defines a new scale that we call $\lambda_{ampl,ij}$, defined to be
\begin{equation}
\lambda_{ampl,ij} =  \frac{ 4.8096\,\pi }{ \breve{V}^{\,C}\,\left(|U_{ei}|^2- |U_{ej}|^2\right) }.
\label{eq:lambdaampl}
\end{equation}
If the combination of the amplitude and wavelength of \emph{any} Fourier mode in the turbulence is such that $G_{a}\,\lambda_{a} \geq \lambda_{ampl,ij}$, then this Fourier mode will have a value for $z$ which places it beyond the first zero of $J_0$. It is easiest to satisfy this condition for larger wavelengths, which typically also have larger amplitudes. Thus we expect the long wavelength modes, particularly those for which $\lambda_{a} \approx \lambda_{cut}$, are most likely to cause suppression. From the definition of $\lambda_{ampl,ij}$, it is clear $\lambda_{ampl,ij}$ decreases as the potential $\breve{V}^{\,C}$ increases, meaning the suppression effect is more important in dense matter. Only close to a MSW resonance, when $|U_{ei}|^2- |U_{ej}|^2$ becomes small, can the effect of large potential be compensated.

\subsection{$h^{\,C}$}

Finally there is the potential scale height $h^{\,C}$, which measures the distance over which the potential changes. This quantity is defined to be
\begin{equation}
h^{\,C} = \frac{\breve{V}^{\,C}}{d\breve{V}^{\,C}/dr}.
\end{equation}
As one observes in figure (\ref{fig:rhoYe}), the gradient of the mass density is very large close to the proto-neutron star, which tends to make the potential scale height small. At larger radii, but still behind the shock, the gradient softens and can even approach zero. The scale height is important in determining the degree to which transitions between states occurs: even if a resonance condition exists and the amplitude is not suppressed by the long wavelength modes, if the potential changes too rapidly -- that is, if the scale height is small -- then the system passes through the resonance too quickly to allow the neutrino to make any substantial transition between the states \cite{Patton:2014lza}.  


\subsection{The three criteria for Stimulated Transitions}

Now that we have defined the six different lengthscales, we see three different criteria must be satisfied in order for turbulence to have an effect. The conditions are, for any pair of states $i$ and $j$: 
\begin{itemize}
 \item $\lambda_{diss} \leq \lambda_{split,ij} \leq \lambda_{cut}$,
  \item $\lambda_{trans,ij} / G_{\star,ij} < h^{\,C}$ for the mode which matches the eigenvalue difference,
  \item $\lambda_{ampl,ij} / G_{a} \geq \lambda_{a}$ for all modes.
\end{itemize}
The physical reasoning for these conditions are quite simple. The splitting scale $\lambda_{split,ij}$ must be between the cutoff scale $\lambda_{cut}$ and the dissipation scale $\lambda_{diss}$ in order that there be a Fourier mode in the turbulence that can match the eigenvalue splitting between the states and thus drive a transition between them. The lengthscale over which the neutrino makes transitions between the states, which we estimate to be $\lambda_{trans,ij}/G_{\star,ij}$, must be smaller than the potential scale height $ h^{\,C}$ in order that there be enough `space' to make the transition while the resonance is fulfilled. Finally, we require that all modes satisfy $G_{a}\,\lambda_{a} \leq \lambda_{ampl,ij}$ otherwise transitions are suppressed by long wavelength modes.

Let us see how these scales look for selected snapshots shown in figure (\ref{fig:rhoYe}).
\begin{figure*}[t!] 
\includegraphics[width=0.49\linewidth]{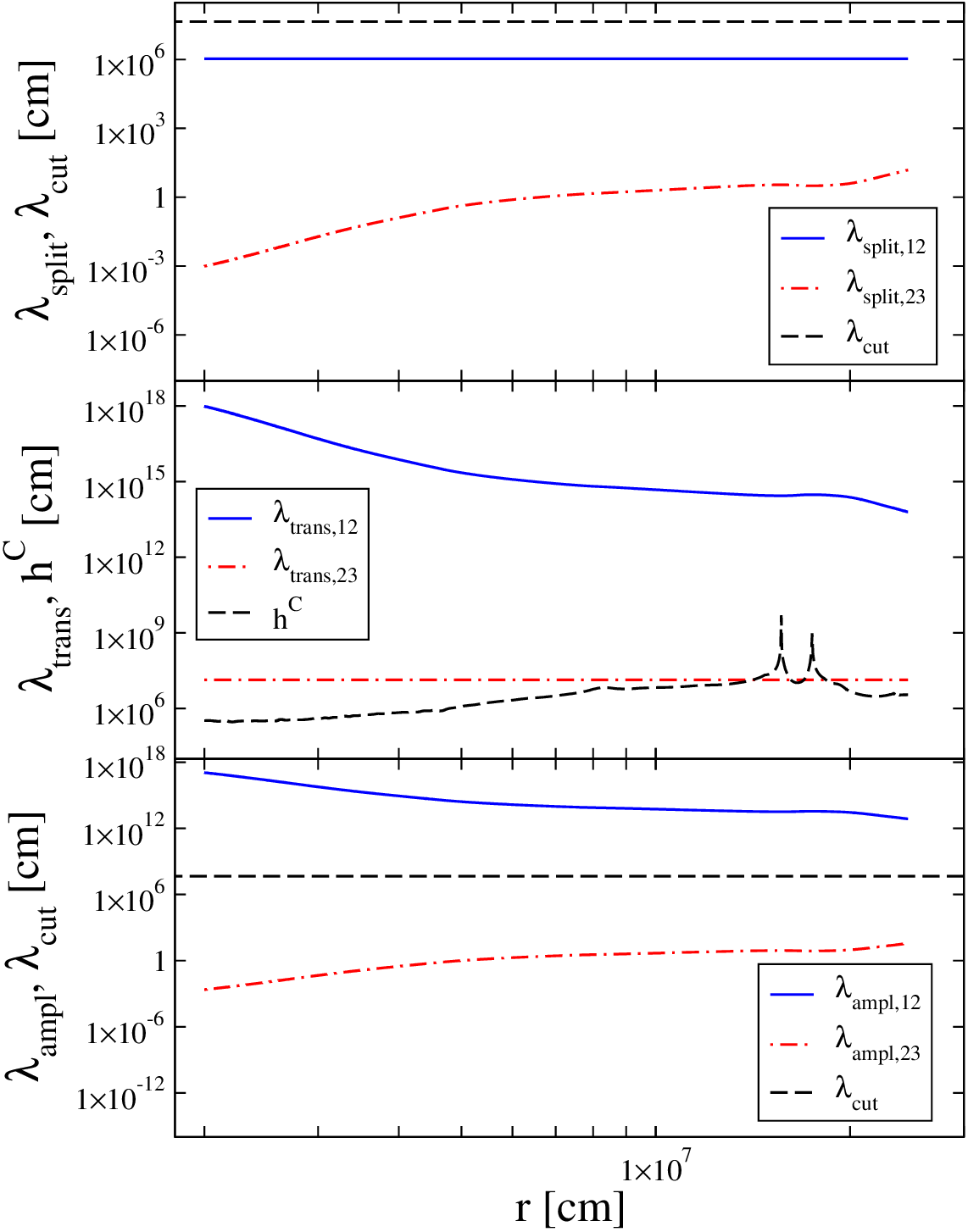}
\includegraphics[width=0.49\linewidth]{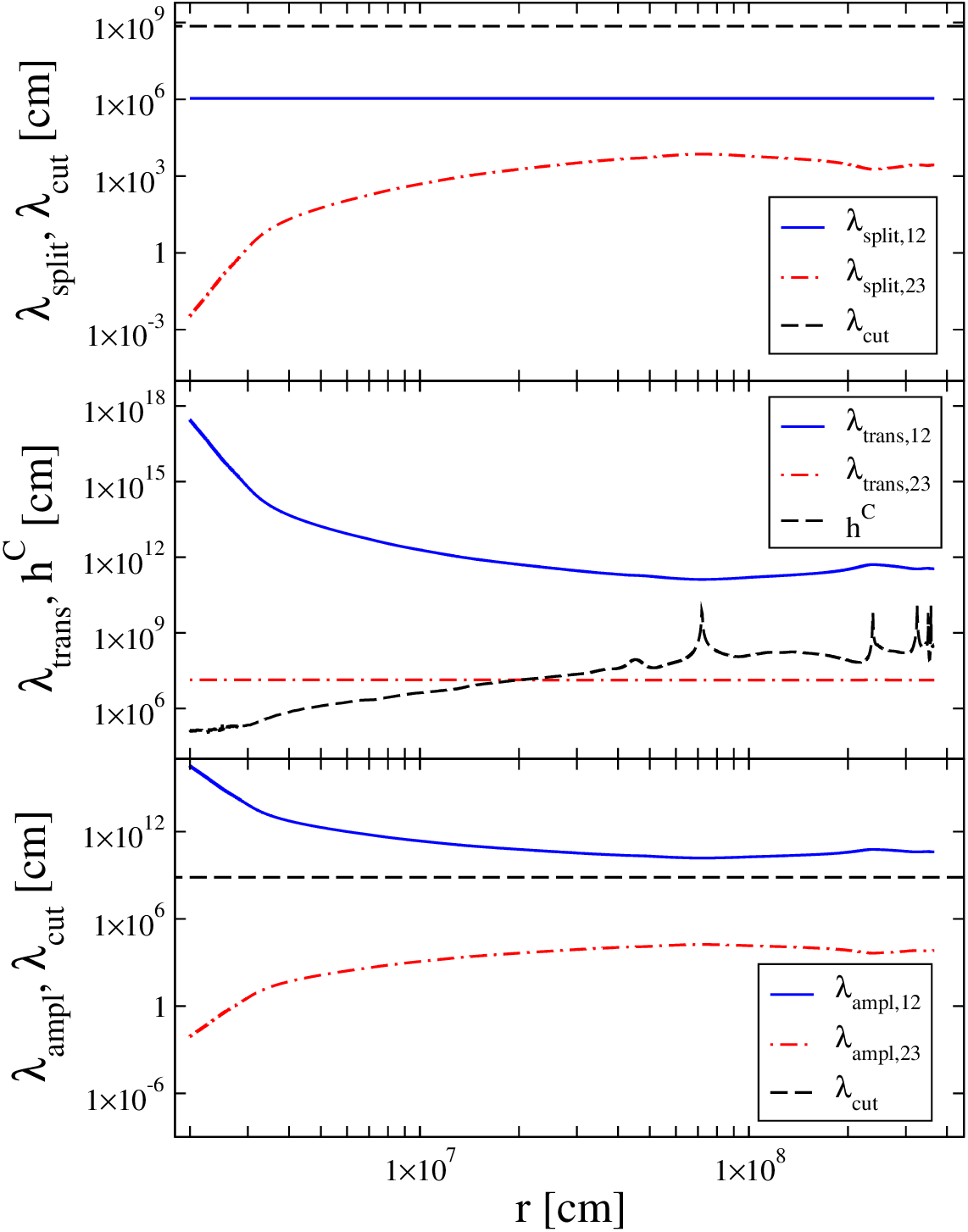}
\caption{The three conditions for turbulence effects for the $t=0.3\;{\rm s}$ and $t=0.7\;{\rm s}$ postbounce snapshot. In the top panels, the scale $\lambda_{cut}$ is the horizontal dashed black line, the splitting scale for states $1$ and $2$ is the blue solid line and between states $2$ and $3$ the red dash-dot line. The middle panels show the scales $\lambda_{trans,ij}$ for states $1$ and $2$ as the blue solid line, between $2$ and $3$ as the red dashed-dot line and the potential scale height $h^{\,C}$ as the black dashed line. In the bottom panel scale $\lambda_{cut}$ is the horizontal dashed black line, the scale $\lambda_{ampl,ij}$ for states $1$ and $2$ is the blue solid line and for states $2$ and $3$ is the red dashed-dot line \label{lambdas3}}
\end{figure*}
The three criteria for the $t=0.3\;{\rm s}$ and $t=0.7\;{\rm s}$ snapshots are shown in figure (\ref{lambdas3}). The top panels show that the first condition, $\lambda_{split,ij} \leq \lambda_{cut}$, is easily satisfied for both the $1-2$ and $2-3$ mixing channels. The cutoff scale, indicated by the horizontal dashed line, is of order $\lambda_{cut} \sim 1000\;{\rm km}$ whereas the splitting between states $1$ and $2$ (solid blue line) corresponds to a wavelength of $\sim 10\;{\rm km}$ for this neutrino energy and mixing parameters, and between states $2$ and $3$ (dash-dotted red line) the splitting scale ranges from $\sim 10\;{\rm \mu m}$ to $\sim 10\;{\rm m}$. The splitting scale between states $1$ and $3$ (not shown) is very close to that between states $2$ and $3$. The splitting scales for the antineutrinos in a normal mass ordering at these densities are such that $\bar{\lambda}_{split,12} \approx \lambda_{split,23}$, $\bar{\lambda}_{split,13} \approx \lambda_{split,13}$ and $\bar{\lambda}_{split,23} \approx \lambda_{split,12}$. The reassignment for an inverted mass ordering have been previously mentioned. Note that the ratio of $\lambda_{cut}$ to $\lambda_{split,ij}$ is the source of the statements found in the Introduction that the resolution of the simulations would need have a dynamical range $\gtrsim 100\;{\rm db}$ if one wanted to account for neutrino flavor transformation due to turbulence. The scale $\lambda_{diss}$ is not shown in either panel but the requirement $\lambda_{diss} \leq \lambda_{split,ij}$ would be satisfied for any dissipation lengthscale of order $\lambda_{diss} \lesssim 10\;{\rm \mu m}$. Recall our estimate was $\lambda_{diss} \approx 10^{-5}\;{\rm cm}$. Even if this was a bad estimate and the dissipation scale were as large as $\lambda_{diss} \sim 1\;{\rm cm}$, the criterion that $\lambda_{diss} \leq \lambda_{split,ij}$ would only mean no turbulence effects could occur for $r \lesssim 50\;{\rm km}$ at $t=0.3\;{\rm s}$ snapshot and $r \lesssim 30\;{\rm km}$ at $t=0.7\;{\rm s}$. A large region where turbulence effects could still occur would remain. Thus we see the first criterion is easily satisfied so we turn our attention to the second.

In the middle panels we compare $\lambda_{trans,ij}$ and $h^{\,C}$. The transition scale $\lambda_{trans,ij}$, indicated by the blue solid line for $1-2$ and the red dash-dotted line for $2-3$, is the distance over which a neutrino transitions between pairs of states on resonance when the resonant mode has unit amplitude. The transition wavelength between states $\nu_1$ and $\nu_3$ is not shown but is like $\lambda_{trans,23}$ in that it is a constant but at a value approximately an order of magnitude larger. Similarly the transition wavelengths for the antineutrinos in a normal mass ordering  at these densities are such that $\bar{\lambda}_{trans,12} \approx \lambda_{trans,13}$, $\bar{\lambda}_{trans,13} \approx \lambda_{trans,23}$ and $\bar{\lambda}_{trans,23} \approx \lambda_{trans,12}$. If the resonant mode has an amplitude $G_{\star,ij}$, then the wavelength $\lambda_{Q,ij}$ has a lower limit of $\lambda_{trans,ij}/G_{\star,ij}$. For $G_{\star,ij} \leq 1$ we shift the red and blue curves upwards by $1/G_{\star,ij}$; as a result, the regions where the red and blue lines are below the black dashed line corresponding to the potential scale height $h^{\,C}$, become narrower or disappear. The rms amplitude of the resonant mode is a function of the power spectrum. If we do not wish to impose any form for the power spectrum, then we must determine whether there is any value of $G_{\star,ij}$ such that the condition could be satisfied. The figure indicates that no value of $G_{\star,12}$ less than unity could be found at either snapshot for the mixing between states $1$ and $2$, except momentarily at those locations where the potential scale height becomes infinite due to a potential minimum. So for mixing between states $1$ and $2$, the second condition cannot be satisfied and we do not expect turbulence effects in this channel. 

The mixing channel between states $2$ and $3$ looks more promising. We can find a region around $r\sim 150\;{\rm km}$ at $t=0.3\;{\rm s}$ where $\lambda_{trans,ij} < h^{\,C}$ and at $t=0.7\;{\rm s}$ the entire profile from $r\sim 150\;{\rm km}$ to the shock satisfies $\lambda_{trans,ij} < h^{\,C}$. But the actual condition we need to satisfy is $\lambda_{trans,23}/G_{\star,23} < h^{\,C}$ so for the $t=0.3\;{\rm s}$ snapshot the amplitude of the resonant mode would have to be very large in order to satisfy $\lambda_{trans,23}/G_{\star,23} < h^{\,C}$. The top panel indicates the wavelength of this mode, which is equal to $\lambda_{split,23}$ for the $2-3$ channel, would be of order $\lambda_a \approx 1\;{\rm cm}$ at $r\sim 150\;{\rm km}$. This is much smaller than the cutoff scale, $\lambda_{cut} \sim 1000\;{\rm km}$, so the power spectrum would need to be very hard. At $t=0.7\;{\rm s}$ the amplitude of the resonant modes would not need to be as large because $\lambda_{trans,23} < h^{\,C}$ by 1-2 orders of magnitude for $r\gtrsim 150\;{\rm km}$. At the same time, the wavelength of the resonant mode has increased to $\lambda_a \approx 1000\;{\rm cm}$ as shown in the top panel. Thus we could accommodate a resonant amplitude of order $G_{\star,23} \sim 10^{-2}$ and still satisfy the condition $\lambda_{trans,23}/G_{\star,23} < h^{\,C}$. This amplitude does not require as hard a power spectrum as that required at $t=0.3\;{\rm s}$ because the scale is closer to the cutoff scale. We therefore find that the second criterion for turbulence effects in the $2$-$3$ channel can be satisfied but not as easily as the first; in order to have an effect, the turbulence requires large amplitudes and hard power spectra. Even the hard turbulence power spectra found by Abdikamalov \emph{et al.} \cite{2015ApJ...808...70A} may not be able to produce the amplitudes required. Given that $\lambda_{trans,13}$ is similar to $\lambda_{trans,23}$ but an order of magnitude larger, large amplitudes are also required for turbulence effects between states $1$ and $3$.

Finally, we compare $\lambda_{ampl,ij}$ and $\lambda_{cut}$ in the bottom panels. For the sake of clarity the amplitude suppression wavelength for states $\nu_1$ and $\nu_3$ is not shown but is found to be identical to $\lambda_{ampl,23}$ at these densities. Similarly the suppression wavelengths for the antineutrinos in a normal mass ordering at these densities are such that $\bar{\lambda}_{ampl,12} \approx \lambda_{ampl,23}$, $\bar{\lambda}_{ampl,13} \approx \lambda_{ampl,13}$ and $\bar{\lambda}_{ampl,23} \approx \lambda_{ampl,12}$. The requirement that $\lambda_{ampl,ij} / \lambda_a \geq G_a \forall a$ in order to see turbulence effects is hardest to satisfy for the longest wavelength modes, which have wavelengths of order $\lambda_{cut}$. For any other mode, $\lambda_a \leq \lambda_{cut}$, so the maximum amplitude of mode $a$ before suppression occurs will be larger than $\lambda_{ampl,ij} / \lambda_{cut}$. Though already ruled out by the second criteria, the bottom panels indicate that the ratio of $\lambda_{ampl,12} / \lambda_{cut}$ is much larger than unity for the $1-2$ mixing channel. There appears to be no amplitude suppression effect for states $1$ and $2$. In contrast, the ratio $\lambda_{ampl,23} / \lambda_{cut}$ is of order $\sim 10^{-6}$ for the $2-3$ mixing channel, meaning the amplitude of the longest wavelength modes would need to be $\lesssim 10^{-6}$ in order to see turbulence effects. This condition is completely at odds with the conclusion from our analysis of the middle panels, where we found we required $G_{\star,23} \gtrsim 10^{-2}$ in order to make $\lambda_{trans,23}/G_{\star,23} < h^{\,C}$. The region of the turbulence parameter space that was not ruled out after our analysis of the second criteria has now been eliminated by trying to satisfy the third. Given the similarity between  $\lambda_{ampl,13}$ and $\lambda_{ampl,23}$, turbulence effects in the $1$-$3$ channel are also ruled out.

Thus we find it appears impossible to satisfy the three conditions simultaneously in the $1-2$, $1-3$ and $2-3$ mixing channels: the turbulence amplitude must, contrarily, be large in order to satisfy $\lambda_{trans,ij}/G_{\star,ij} < h^{\,C}$, while also being small so that $\lambda_{ampl,ij} / G_{a} \geq \lambda_{cut}$. We expect no effect from turbulence due to Stimulated Transitions. This result is not sensitive to the neutrino energy for typical supernova energies of $1\;{\rm MeV}$ to $100\;{\rm MeV}$. This is because for a normal mass ordering, the eigenvalue $k_3$ and the element $U_{e3}$ of the mixing matrix $U$ approach asymptotes at high density which are independent of the neutrino energy \cite{2009PhRvD..80e3002K} though energy dependence remains in $k_1$, $k_2$, $U_{e1}$ and $U_{e2}$. The effect of the asymptotes is that $\lambda_{split,13}$, $\lambda_{split,23}$, $\lambda_{ampl,13}$ and $\lambda_{ampl,23}$ are all independent of energy at these densities, $\lambda_{split,12}$, $\lambda_{trans,13}$ and $\lambda_{trans,23}$ are linearly proportional to the energy, and $\lambda_{trans,12}$ and $\lambda_{ampl,12}$ are quadratically proportional to the energy. Changing the neutrino energy does not make it easier to satisfy the three criteria. For example: in order to make it easier to satisfy the criterion $\lambda_{trans,23} / G_{\star,23} < h^{\,C}$ in the $\nu_2 - \nu_3$ mixing channel one could lower the neutrino energy, except the third criterion, $\lambda_{ampl,23} / G_{a} \geq \lambda_{a}$, would not change and already rules out a turbulence effect. The only strategy that could satisfy the criteria is to lower the neutrino energy dramatically until the MSW densities approach the densities of the matter in the turbulence region at which point the scaling with neutrino energy is no longer valid. This requires the neutrino energy be lower than $E \lesssim 10\;{\rm keV}$ for the $2-3$ mixing channel and $E \lesssim 100\;{\rm eV}$ for the $1-2$ channel. 


\section{Distorted phase effects}
\label{sec:distortedphase}

The second path by which turbulence can affect neutrino propagation is known as Distorted Phase Effects \cite{2015PhRvD..92a3009K}. The theory behind the effect is much simpler than the Stimulated Transitions model, and the criteria for determining whether an effect occurs is also much simpler to understand. Distorted Phase Effects for neutrinos are very similar to what one expects in the intensity of light reflected from a thin film with a spatially-varying index of refraction. The effect relies upon multiple semi-adiabatic MSW resonances or discontinuities in the density profile where the neutrinos do not completely swap flavor as they cross. The neutrino evolution is adiabatic before, between, and after these semi-adiabatic MSW resonances or discontinuities.

Let us consider a neutrino propagating from some initial point $r_0$ to $r$ through two discontinuities located at positions $r_1$ and $r_2$. In the unperturbed matter basis, the evolution matrix from $r_0$ to $r$ can be broken down into the evolution from $r_0$ to $r_1$, the evolution across the discontinuity at $r_1$, the evolution from $r_1$ to $r_2$, the evolution across the discontinuity at $r_2$, and finally the evolution from $r_2$ to $r$. That is, $S(r,r_0) = S(r,r_2+)\,S(r_{2+},r_{2-})\,S(r_2-,r_1+)\,S(r_{1+},r_{1-})\,S(r_{1-},r_0)$. The evolution matrix describing the evolution through a discontinuity can be written as 
\begin{equation}
S(r_+,r_-) = U^{\dagger}(r_+)\,U(r_-) =  \left( \begin{array}{ccc} \sqrt{P_{11}}\,e^{\rmi\, \chi_{11}} & \sqrt{P_{12}}\,e^{\rmi\, \chi_{12}} & \sqrt{P_{13}}\,e^{\rmi\, \chi_{13}} \\ 
\sqrt{P_{21}}\,e^{\rmi\, \chi_{21}} & \sqrt{P_{22}}\,e^{\rmi\, \chi_{22}} & \sqrt{P_{23}}\,e^{\rmi\, \chi_{23}} \\ 
\sqrt{P_{31}}\,e^{\rmi\, \chi_{31}} & \sqrt{P_{32}}\,e^{\rmi\, \chi_{32}} & \sqrt{P_{33}}\,e^{\rmi\, \chi_{33}}  \end{array} \right)
\end{equation}
with $P_{ij}$ the transition probabilities for the matter basis states across the discontinuity and the $\chi_{ij}$'s are a set of phases. The evolution matrices for the adiabatic sections are given by 
\begin{equation}
S(r_2,r_1) = \exp\left( -\rmi \int_{r_1}^{r_2}\,dr \left\{ K(r) + \Xi(r) \right\} \right) = \exp\left( -\rmi\, \Phi(r_2,r_1) \right) \label{eq:adiabaticS}
\end{equation}
where $K$ is the matrix of eigenvalues of the unperturbed Hamiltonian and $\Xi$ was the matrix that canceled the diagonal element of the Hamiltonian for the $B$ matrix in equation (\ref{dBdt}). This equation for $S(r_2,r_1)$ comes from solving equation (\ref{eq:dbreveSdt}) assuming $U^{\dagger} dU/dr$ is negligible, the definition of the $W$ matrix and using $B(r_2,r_1)=1$. Since $K$ and $\Xi$ are both diagonal, the matrix $\Phi(r_2,r_1)$ in equation (\ref{eq:adiabaticS}) is also a diagonal matrix $\Phi=\rm{diag}( \phi_1, \phi_2, \ldots)$. Putting the pieces together and computing the transition probability between states $i$ and $j$, we find 
\begin{equation}
P_{ij} = \sum_{k} P_{ik}(r_2)\,P_{kj}(r_1) 
+ 2\sum_{k,l>k}\sqrt{P_{ik}(r_2)\,P_{kj}(r_1)\,P_{i\ell}(r_2)P_{\ell j}(r_1)}\,\cos \psi 
\label{eq:PijDPE}
\end{equation}
with $\psi = \phi_\ell(r_2,r_1) -\phi_k(r_2,r_1) + \chi_{i\ell}(r_2)-\chi_{ik}(r_2)+\chi_{\ell j}(r_1)-\chi_{kj}(r_1)$. The second term in equation (\ref{eq:PijDPE}) is an interference term and depends upon the phases $\phi(r_2,r_1)$ which, as equation (\ref{eq:adiabaticS}) shows, depend upon the turbulence because $\Xi$ is non-zero when the turbulence is present. Note that the interference term in the expression for the transition probability $P_{ij}$ does not exist in the region between the first and second discontinuity, it appears only after the second discontinuity is passed. 

There are many variations on this same basic phenomenon, but in all cases one observes the effect of turbulence due to a change in the phase of the neutrino, not because of any transitions induced between the states in the region where the turbulence occurs. The important feature in all of them is the presence of the discontinuities where the phase of the neutrino becomes important. Discontinuities can be found by computing the diabaticity $\Gamma_{ij}$ between states $i$ and $j$, defined\footnote{Note we have ignored the term $\delta Q_{ij}$ from \cite{2012JPhG...39c5201G} because the off-diagonal element of the unperturbed Hamiltonian has no imaginary component} to be \cite{2012JPhG...39c5201G}
\begin{equation}
\Gamma_{ij} = -\frac{2\pi}{\delta k_{ij} }\left( U^{\dagger}\,\frac{dU}{dr} \right)_{ij}.
\end{equation}
Again $k_i$ are the eigenvalues of the Hamiltonian and the term in parentheses in this equation can be shown to be
\begin{equation}
\left( U^{\dagger}\,\frac{dU}{dr} \right)_{ij} = -\frac{1}{\delta k_{ij} }\left( U^{\dagger}\,\frac{d\breve{H}^{(f)} }{dr} U \right)_{ij}.
\end{equation}
The diabaticity is related to the lengthscales we introduced in section \S\ref{sec:thescales} because one finds that the $\Gamma_{ij}$ can be written as 
\begin{equation} 
\Gamma_{ij} = 2 \frac{\lambda_{split,ij}^2}{h^C\,\lambda_{trans,ij}}
\end{equation}
Discontinuities or semi-adiabatic MSW resonances appear as locations where $\Gamma_{ij}$ is greater than unity i.e.\ locations where 
\begin{equation}
2 \frac{\lambda_{split,ij}^2}{h^C\,\lambda_{trans,ij}} \geq 1.
\end{equation}
If we find two (or more) locations where $\Gamma_{ij}$ is greater than unity then we have circumstances where Distorted Phase Effects can occur.

\begin{figure*}[t!] 
\includegraphics[width=0.48\linewidth]{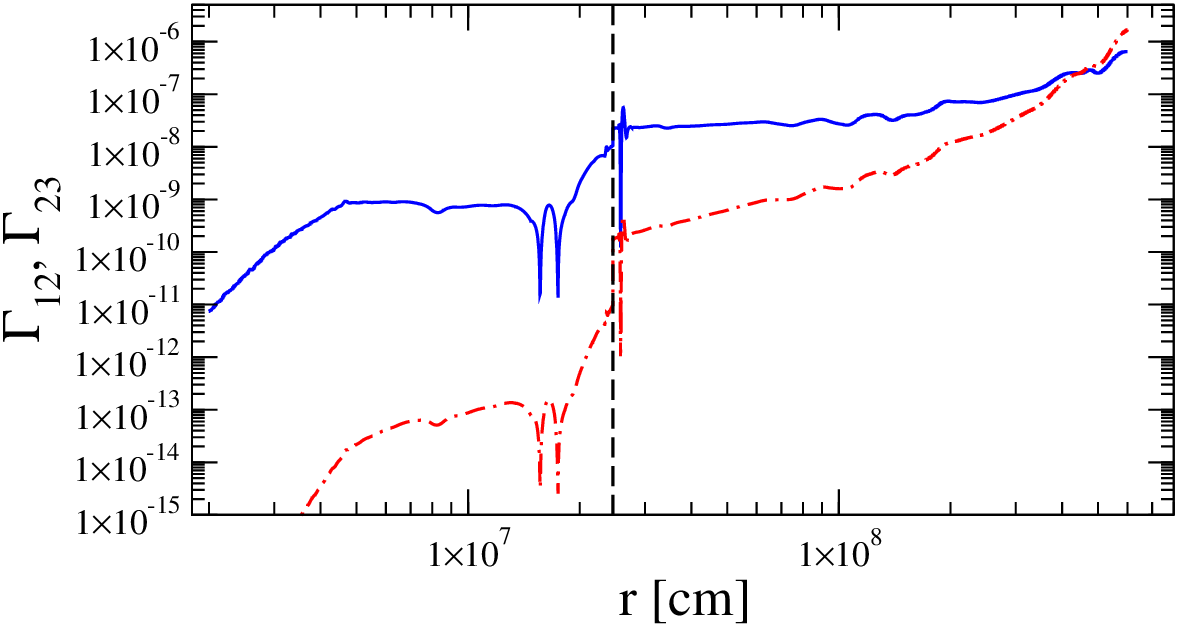}
\includegraphics[width=0.48\linewidth]{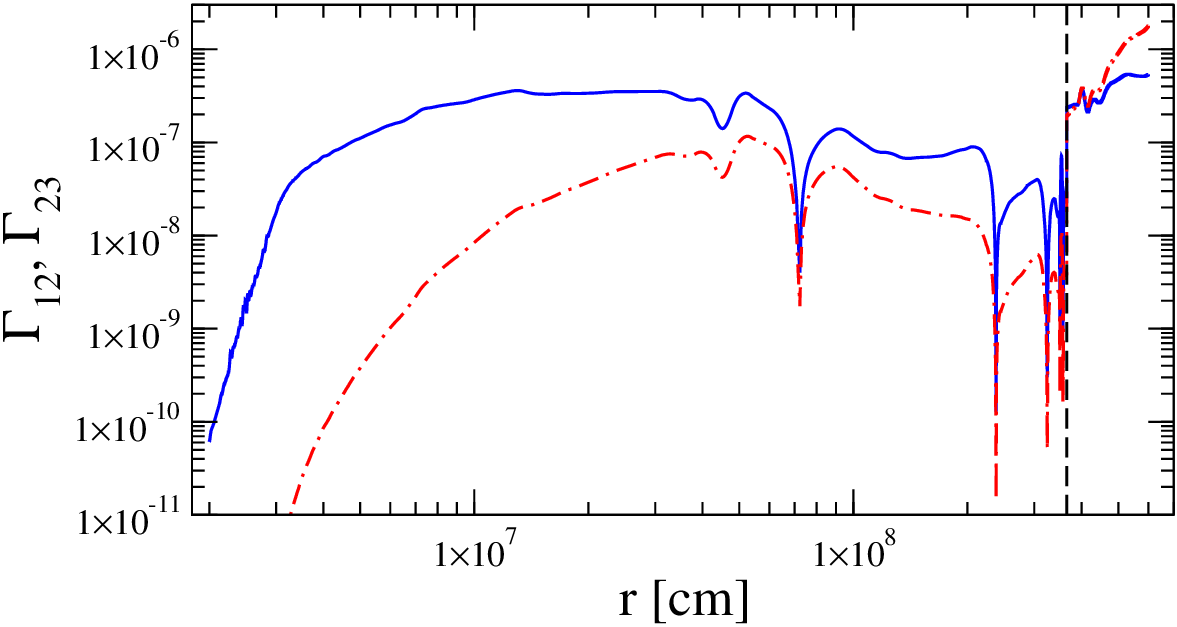}
\caption{The diabaticity parameters $\Gamma_{12}$ (blue solid line) and $\Gamma_{23}$ (red dot-dash line) as functions of distance $r$ for two snapshot times. On the left $t=0.3\;{\rm s}$ and on the right $t=0.7\;{\rm s}$. 
The position of shock at these times is indicated by the dashed vertical line. \label{Gammas3} }
\end{figure*}
The diabaticities for the $t=0.3\;{\rm s}$ and $t=0.7\;{\rm s}$ snapshots are shown in figure (\ref{Gammas3}). To make these plots we have used the same masses, mixing angles and neutrino energy as for the numerical calculations. For the sake of clarity the diabaticity between states $\nu_1$ and $\nu_3$ is not shown but is found to be smaller than $\Gamma_{23}$ by approximately an order of magnitude. The diabaticity of the antineutrinos, denoted by $\bar{\Gamma}_{ij}$, in a normal mass ordering at these densities are found to be $\bar{\Gamma}_{12} \approx \Gamma_{13}$, $\bar{\Gamma}_{13} \approx \Gamma_{23}$ and $\bar{\Gamma}_{23} \approx \Gamma_{12}$. The position of the shock at these times is indicated by the vertical dashed line. The reader will observe there are no locations before the shock where $\Gamma_{12}$ or $\Gamma_{23}$ approach unity. Thus one does not expect Distorted Phase Effects due to turbulence: the semi-adiabatic MSW resonances and discontinuities we require to generate the effect are not present. Again this result is not sensitive to the neutrino energy for energies typically found in supernovae. Using the scalings of $\lambda_{split,ij}$ and $\lambda_{trans,ij}$ given in section \S\ref{sec:thescales} we find $\Gamma_{12}$ is independent of energy while $\Gamma_{13}$ and $\Gamma_{23}$ are both inversely proportional to the neutrino energy. Thus higher energy neutrinos are more diabatic than lower energies but, given the results in figure (\ref{Gammas3}), we would need to raise the neutrino energy to $E \gtrsim 1\;{\rm TeV}$ in order to achieve $\Gamma_{13} \gtrsim 1$ or $\Gamma_{23} \gtrsim 1$.


\section{Discussion and Conclusions}
\label{sec:Conclusions}

Turbulence during the accretion phase of a core-collapse supernova has been shown to affect the dynamics of the explosion. From previous knowledge of the effect of turbulence upon neutrinos, one can make a plausible argument that accretion phase turbulence could affect the flavor evolution of the neutrinos passing through it. Such changes would then alter the neutrino heating in the gain region and therefore also change the dynamics of the explosion. In this paper we have studied whether turbulence is expected to have this additional indirect effect upon the supernova dynamics, and found that it cannot. Neither mechanism by which turbulence affects neutrinos is operating during the accretion phase in the location of the turbulence. 

For the direct stimulation of transitions between the neutrino states, we showed that three criteria need to be satisfied if turbulence is to change the neutrino flavor evolution. These three conditions are between the six lengthscales that describe the phenomenon: the turbulence cutoff scale $\lambda_{cut}$, the turbulence dissipation scale $\lambda_{diss}$, the density scale height $h^{\,C}$, the splitting scale $\lambda_{split,ij}$ for a given pair of neutrino states, the transition wavelength $\lambda_{trans,ij}$, and finally the amplitude suppression wavelength $\lambda_{ampl,ij}$. The first condition, $\lambda_{diss} \leq \lambda_{split,ij} \leq \lambda_{cut}$, was easily met in all mixing channels. The second condition, $\lambda_{trans,ij} / G_{\star,ij} < h^{\,C}$ for the mode which matches the eigenvalue difference between two neutrino states with amplitude $G_{\star,ij}$, was not possible to satisfy for mixing between states $1$ and $2$ but could be satisfied for the mixing between states $2$ and $3$ and states $1$ and $3$if the amplitude of the resonant mode was large. However, the third condition, $\lambda_{ampl,ij} / G_{a} \geq \lambda_{a}$ for all modes $a$, could not be satisfied simultaneously for neither the $2-3$ nor the $1-3$ mixing channel because it requires the longest wavelength modes have small amplitudes. 

We also found that the second mechanism by which turbulence affects neutrinos, Distorted Phase Effects, is also not in operation during the accretion phase epoch. The multiple semi-adiabatic MSW resonances and/or discontinuities that are required for this effect to occur are simply not present for the neutrino energies of interest. 

Thus based on our analytic models, we do not expect to see turbulence effects during the accretion phase of a core-collapse supernova. This conclusion is not sensitive to the neutrino energy for energies typically found in supernovae and also insensitive to the mass ordering of the neutrinos. We also expect this conclusion to be independent of the progenitor. Comparing the simulations of the $10.8\;{\rm M_{\odot}}$ and $18\;{\rm M_{\odot}}$ progenitors in Fischer \emph{et al.} \cite{2010A&A...517A..80F}, we observe the mass densities below the shock during the accretion phase - which lasts for $\sim 350 \;{\rm ms}$ for both - are similar. The accretion phase of the $8.8\;{\rm M_{\odot}}$ ONeMg progenitor by Fischer \emph{et al.} lasts only $\sim 30\;{\rm ms}$ but during this brief phase, the density behind the shock is still $\rho \gtrsim 10^{7}\;{\rm g/cm^3}$ which is similar to the $t=0.3\;{\rm s}$ profile shown in figure (\ref{fig:rhoYe}). One also might question whether significant turbulence develops for this particular case. 

Finally, it is also consistent with the results from our numerical calculations shown in figures (\ref{fig:Pvsr465}), (\ref{fig:Pvsr492}) and (\ref{fig:Pvsr509}) and the conclusion of the study by Reid, Adams \& Seunarine \cite{2011PhRvD..84h5023R}. 

Our conclusion is not in conflict with the numerous studies which indicate turbulence can have an effect during the cooling phase. During the cooling phase, the turbulence moves out into the supernova to lower densities at or below the MSW resonances. At lower densities the amplitude suppression scale $\lambda_{ampl,ij}$, given by equation (\ref{eq:lambdaampl}), the transition scale $\lambda_{trans,ij}$, given in equation (\ref{eq:lambdatrans}), and the splitting scale $\lambda_{split,ij}$ have all grown considerably and much more than the cutoff scale $\lambda_{cut}$. At the same time, the density profile typically becomes shallower as the star explodes, leading to larger values of the density scale height $h^{\,C}$. All these changes in the lengthscales facilitate the emergence of turbulence effects in the neutrinos. The larger value of $\lambda_{split,ij}$ compared to $\lambda_{cut}$ raises the amplitude of the resonant mode $G_{\star,ij}$. The transition wavelength $\lambda_{trans,ij}$ is also longer, but the second criterion that $\lambda_{trans,ij}/G_{a} < h^{\,C}$ becomes easier to satisfy; most importantly, the third criterion that $\lambda_{ampl,ij}/G_a > \lambda_{cut}$ for all turbulence modes becomes much easier to satisfy simultaneously with the second requirement. 

Finally, one important effect we have not included is neutrino self-interaction. This is a reasonable assumption given the results of Chakraborty \emph{et al.} \cite{2011PhRvL.107o1101C}, who found that the mass density close to the proto-neutron star during the accretion phase was so high that self-interactions were suppressed. Reid, Adams \& Seunarine \cite{2011PhRvD..84h5023R} did not see large self-interaction effects from turbulence in their `single-angle' calculations. However, it remains an open question whether turbulence during the accretion phase may have an effect via self-interaction if so-called multi-angle calculations \cite{Duan:2006an,Duan:2006jv} were undertaken.


\ack

This research was supported by DOE awards DE-SC0006417 and DE-FG02-10ER41577. 
The authors wish to thank Evan O'Connor for providing the data used in deriving the fits to the anisotropic velocity, and to Davide Radice for useful discussions concerning the dissipation lengthscale in supernova turbulence. \\


\bibliographystyle{IEEEtran}
\bibliography{accretion}

\end{document}